\documentclass[12pt]{article}
\usepackage{epsfig}
\begin{document}
\begin{center}
{\Large \bf Microscopic theory of the two-dimensional quantum \\
antiferromagnet in a paramagnetic phase}

{\large V. I. Belinicher$^{*\dagger}$, and J. da Providencia$^{*}$}
\end{center}

\noindent
$^{*}$ University of Coimbra, 3000, Coimbra, Portugal \\
$^{\dagger}$Institute of Semiconductor Physics, 630090, Novosibirsk,
Russia

\begin{center}
{\Large \bf Abstract}
\end{center}

 We have developed a consistent theory of the Heisenberg quantum
antiferromagnet in the disordered phase with a short range
antiferromagnetic order on the basis of the path integral for spin
coherent states. We have presented the Lagrangian of the theory in
a form which is explicitly invariant under rotations and found natural
variables in terms of which one can construct a natural perturbation
theory. The short wave spin  fluctuations are similar to those in  the
spin wave theory and they are of the order of the parameter $1/2s$
where $s$ is the spin magnitude.  The long wave spin fluctuations are
governed by the nonlinear sigma model and are of the order of the
the parameter $1/N$, where $N$ is the number of field
components. We also have shown that the short wave spin fluctuations
must be evaluated accurately and the continuum limit in time of the
path integral must be performed after all summation over the frequencies
$\omega$.
In the framework of our approach we have obtained the response function for the
spin fluctuations for all region of the frequency $\omega$ and the wave
vector ${\bf k}$ and have calculated the free energy of the system.
We have also reproduced the known results for the spin correlation length in
the lowest order in $1/N$.

\vskip 0.5cm
\noindent
Pacs: 75.50.Ee,74.20.Mn
%
\section{Introduction}
\label{sec:I1}
The theory of the two--dimensional Heisenberg antiferromagnet (AF) has
attracted great interest during the last years in connection
with the problem of AF fluctuations in the
copper oxides 
\cite{Man1,Cha1,Sac1,Pi1}.
We especially call the attention of the reader to the
review \cite{Man1} in which a general situation of the quantum AF (QAF)
has been elucidated. The approach of these papers was based
on the sigma model, which describes the long
wave fluctuations of the Heisenberg AF in the paramagnetic phase with
a short range antiferromagnetic order \cite{Sac2}. The sigma model
is the continuous
model for the unit vector ${\bf n}(t,{\bf r}), \ {\bf n}^2 =1$ in the 1 +
2 time and space dimensions \cite{Pol1,Zin1}. As a long wave theory the
sigma model can make a lot of physical predictions such as the
structure of the long wave fluctuations and the magnitude of the
correlation length \cite{Cha1,Sac1,Has1}. The basic parameters of the
sigma model, the spin rigidity $\rho_s$, the velocity of sound $c_s$,
and the perpendicular susceptibility $\chi_{\perp}=\rho_s/c_s^2$
are calculated independently in the framework of the spin wave theory
\cite{Iga1}.

 The derivation of the sigma model presented in the papers
\cite{Man1} and \cite{Sac2}, without taking into account
fluctuations, is equivalent to the statement that the  classical equations
of motion for the spin fluctuations in the QAF are described by the
sigma model.  But up to now a consistent theory of the spin fluctuations
for the QAF with short range AF order was absent. Such is precisely the
topic of this paper.

Our approach to the description of the QAF is based on the
functional integral for the generalized partition function (GPF) in terms
of  spin coherent states. This approach permits to solve some
problems of the theory.

  This paper contains the following basic results:

1) A method of constructing
spin coherent states (Appendix \ref{sec:apA})  invariant under rotations
was proposed. It permits to
write out the Berry phase and its generalization for the final time step
the Perelomov phase \cite{Per1} for the QAF in a form which is explicitly
invariant under rotations.  By the way the Lagrangian for QAF is explicitly
invariant under rotations which permits to construct the theory of QAF with
short range AF order in an invariant manner 
(section \ref{sec:d22}).

2) Variables $\mbox{\boldmath{$\Omega$}}$ and ${\bf M},$
which describes AF and ferromagnetic spin fluctuations,
have been introduced . These fields
$\mbox{\boldmath{$\Omega$}}$ and ${\bf M}$ are not free. Inded
$\mbox{\boldmath{$\Omega$}}^2=1$ and $\mbox{\boldmath{$\Omega$}}\cdot
{\bf M}=0$. We remove the first constraint with the help of the  well
known method of the Lagrange multiplier $\lambda$.  For the elimination of
the second constraint we have used the Faddev - Popov trick which
previously was used in quantum electrodynamics (section \ref{sec:gauge}).

3) We have  formulated he basic $1/2s$ approximation (section \ref{sec:d23}) in
the framework of which we have calculated the basic objects of the theory:
a) the Green functions of the $\mbox{\boldmath{$\Omega$}}$, ${\bf M}$,
and $\lambda$ fields; b) 
the correlation length in a zero
approximation; c) the spin correlation function as function of the
frequency $\omega$ and the wave vector ${\bf k}$ (section \ref{sec:spic})

4) In the leading approximation in $1/2s$ we have integrated the action over
the ${\bf M}$ field and we have obtained (section \ref{sec:sprot}) some sort of the
quantum lattice rotator model (QLRM) \cite{Man1}, \cite{Cha1}. We shall
call this model the spin--rotator model. It can be useful in the
calculations in the basic $1/2s$ approximation.

5) It was demonstrated that the continuous approach in time can not be
applied for calculations of the corrections to the basic approximation.
The free energy also can not be calculated on the basis of the continuous
approach in time.
The method of calculations at the final time step was developed
(section \ref{sec:bas3}). After that we calculated the free energy
(section \ref{sec:fren}), and the first order corrections to the basic
approximation (section \ref{sec:pert}): the effective action, the magnon
dispersion law, and the correlation length.

6) We have performed the separation of fluctuations for the
$\mbox{\boldmath{$\Omega$}}$ field by introducing some separation scale
$\hbar/\xi \ll \Lambda \ll \hbar\pi /a$ in momentum space.
This separation of scales was performed on the basis of the Pauli -
Villars transformation.
As a result, we have obtain the long wave nonlinear sigma model with additional
contributions from short wave fluctuations. We presented an arguments showing
that these short wave contributions can be included in the renormalization
of the spin stiffness and the velocity of sound.

\section{Basic continuous approximation}
\label{sec:d2}
\subsection{Magnetic fluctuations and 1/2s approximation}
\label{sec:d21}
We consider the spin system which is described by the following
Heisenberg Hamiltonian:
\begin{eqnarray} \label{2.1}
\hat{H}_{Hei} = \frac{J}{2}\sum_{l,l'=<l>}\hat{\bf S}_{l} \cdot
\hat{\bf S}_{l'}, \ \ \ \hat{\bf S}_{l}\cdot\hat{\bf S}_{l} = s(s+1),
\end{eqnarray}
where $\hat{\bf S}_{l}$ are the spin operators; the index $l$ runs over
a two--dimensional square lattice; the index $l'$ runs over
the nearest neighbors of the site $l$;
$J>0$ is the exchange
constant which, since it is positive,
corresponds to the AF spin
interaction; and $s$ is the magnitude of spin.
 The most efficient method of dealing with a spin system is based on the
representation of the GPF $Z$ or the
generating functional of the
spin Green functions in the form of the functional
integral over spin coherent states $z,z^*$ \cite{Kla1,Vie1} or over
the unit vector ${\bf n}, \ {\bf n}^2=1,$ on a sphere \cite{Man1}:
\begin{eqnarray}  \label{2.2}
&&Z = Tr\left[\exp\left(-\beta\hat{H}\right)\right], \ \ \
\beta = 1/T,
\\  \label{2.2a}
&&Z = \int_{-\infty}^{\infty} \cdots
\int_{-\infty}^{\infty}
D\mu({\bf n}) \exp(A({\bf n})),
\\  \label{2.2b}
&&D\mu({\bf n})=\prod_{\tau l} \frac{2s+1}{2\pi}
\delta({\bf n}_{\tau l}^2-1)d{\bf n}_{\tau l}
\end{eqnarray}
where $T$ is the temperature, $\tau$ is the imaginary time, and $A({\bf
n})$ is the action of the system. In the continuum approximation,
which is valid in leading order over $1/2s$ the expression of the action
$A({\bf n}) $is simplified
\begin{eqnarray} \label{2.3}
&&A({\bf n}) = -\int_{0}^{\beta}\sum_{l}{\cal L}_{tot}(\tau,l)d\tau, \ \ \
{\cal L}_{tot}(\tau,l)={\cal L}_{kin}(\tau,l)+{\cal H}(\tau,l),
\\ \label{2.3a}
&&{\cal L}_{kin}(\tau,l) = is(1-\cos\theta_{\tau l})
\dot{\varphi}_{\tau l}, \ \ \
{\cal H}(\tau,l) = \frac{Js^2}{2}\sum_{l'=<l>}{\bf n}_{\tau l}
\cdot{\bf n}_{\tau l'},
\end{eqnarray}
where $\theta, \varphi$ are the Euler angles of the unit vector
${\bf n}=(\cos\varphi\sin\theta,\sin\varphi\sin\theta,\cos\theta)$,
and ${\bf \dot{\bf \varphi}}_{\tau l}$ is the time derivative of this
angle. The kinetic part of the action ${\cal L}_{kin}$ is highly nonlinear
and it is not clear how to proceed with it consistently. Some essential
steps in this direction where made in \cite{Man1,Sac2} but they do
not have a final character.

Further in this paper we use the idea of the near AF order.
Following this fundamental hypothesis, we split our square lattice
into two AF sublattices ${\rm a}$ and ${\rm b}$.
In the sublattice ${\rm a}$
the spins ${\bf S}$ are directed along some axis ${\bf \Omega }$, in the
sublattice ${\rm b}$ they are directed in the opposite direction.
In this way, we
obtain a new square lattice with two spins ${\rm a}$ and ${\rm b}$ in the elementary
cell with a volume $2a^2$, where $a$ is the space distance between spins.
The axes of this new lattice are rotated
by 45 degrees with respect to the primary axes.
We assume that this AF order is only defined
locally and any global AF order is absent.
As a result, the summation over the lattice sites $l$ and $l'$ can be
expressed as a summation over $l \in {\rm a}$ and $l' \in {\rm b}$.
which specify the space positions of the spins in the
sublattices ${\rm a}$ and ${\rm b}$. In terms of this notation,  the
Lagrangian ${\cal L}_{kin}$ can be expressed as a sum of two such
Lagrangians, one for the  sublattice ${\rm a}$ and another for the
sublattice ${\rm b}$ in terms of two vectors
${\bf n}_a(\tau,l)$ and ${\bf n}_b(\tau,l')$.  The Hamiltonian ${\cal H}$
conserves its form if $l \in {\rm a}$ and $l' \in {\rm b}$ but 
$J/2 \rightarrow J$ because the double summation is absent now.

 In this way we have two spins in
each AF elementary
cell which are defined in the different space positions
$l$ and $l'$.
This circumstance is not convenient for subsequent
nonlinear changes of variables. One can introduce new variables
${\bf n}_{a,b}(\tau,l)$ which are both
defined at the sublattice ${\rm a}$ (or at the center of the AF
elementary cell).

For that we pass
to the Fourier image ${\bf n}_{a,b}(\tau,{\bf k})$ of the original vectors
${\bf n}_a(\tau,l)$ and ${\bf n}_b(\tau,l')$, where the momentum  vector
${\bf k}$
runs over the AF Brillouin band. We can return to the
space representation and consider the coordinate $\mbox{\boldmath{$\rho$}}$
as continuous variable. As a result we have the following definition:
\begin{eqnarray} \label{2.4}
{\bf n}_{a,b}(\tau,\mbox{\boldmath{$\rho$}}) =
\sqrt{1/N_s}\sum_{{\bf k}}\exp\left(i{\bf k}\cdot
\mbox{\boldmath{$\rho$}}\right) {\bf n}_{a,b}(\tau,{\bf k}),
\end{eqnarray}
where $2N_s$ is the total number of sites in the space lattice.
Of course, we
assume periodic boundary conditions.
We can put the  variable $\mbox{\boldmath{$\rho$}}$ now on the sublattice
${\rm a}$ (or in a center of the AF elementary cell).
One can check that the Lagrangian
${\cal L}_{kin}$ will be the same in terms of the new variables
${\bf n}_{a,b}(\tau,\mbox{\boldmath{$\rho$}})\equiv{\bf n}_{a,b}(\tau,l)$.
By the same manner one can change the measure of integration
(\ref{2.2b}) and write out it in terms of
${\bf n}_{a,b}(\tau,l)$.
The Hamiltonian ${\cal H}$ preserves its simple form in the momentum
representation.
\subsection{Invariant Lagrangian}
\label{sec:d22}

The form of the Lagrangian ${\cal L}_{kin}$ is not invariant under
rotations 
although the physical problem itself is invariant.
The reason for such situation was explained in detail in the Appendix
(\ref{sec:Tpscs}). In the Appendix (\ref{sec:Ics}) an invariant form of
the Lagrangian ${\cal L}_{kin}$ was proposed (\ref{A20}). For two
sublattices ${\rm a}$ and ${\rm b}$ it has a form
\begin{eqnarray} \label{2.4a}
{\cal L}_{kin}=-is{\bf n}_a\cdot[{\bf m}_a\times\dot{\bf m}_a]
-is{\bf n}_b\cdot[{\bf m}_b\times\dot{\bf m}_b].
\end{eqnarray}
Eq.(\ref{2.4a}) for ${\cal L}_{kin}$
is valid for any choice the unit
vectors ${\bf m}_{a,b}$ if they satisfy conditions: ${\bf m}^2_{a,b}=1, \
{\bf n}_{a,b}\cdot{\bf m}_{a,b}=0$. For the problem of the quantum
AF with two sublattices we can choose the following
expression for vectors ${\bf m}_{a,b}$
\begin{eqnarray} \label{2.4b}
{\bf m}_{a}=\frac{{\bf n}_{b}-({\bf n}_{b}\cdot{\bf n}_{a}){\bf n}_{a}}
{\sqrt{1-({\bf n}_{a}\cdot{\bf n}_{b})^2}}, \ \ \
{\bf m}_{b}=\frac{{\bf n}_{a}-({\bf n}_{a}\cdot{\bf n}_{b}){\bf n}_{b}}
{\sqrt{1-({\bf n}_{a}\cdot{\bf n}_{b})^2}}.
\end{eqnarray}
This choice means physically that the vectors ${\bf n}_a,{\bf m}_a$
determines the reference frame for the sublattice ${\rm a}$, and
the vectors ${\bf n}_b,{\bf m}_b$
determines the reference frame for the sublattice ${\rm b}$ (see
Appendix (\ref{sec:Ics})).
Substituting these expressions for ${\bf m}_{a,b}$ into Eq. (\ref{2.4a})
we get for ${\cal L}_{kin}$ an invariant over rotation form for
${\cal L}_{kin}$ and also ${\cal H}$ (see (\ref{2.3a}))
\begin{eqnarray} \label{2.4c}
&&{\cal L}_{kin}=\frac{is}{1-{\bf n}_{a\tau l}\cdot{\bf n}_{b\tau l}}
(\dot{\bf n}_{a\tau l}-\dot{\bf n}_{b\tau l})\cdot
[{\bf n}_{a\tau l}\times{\bf n}_{b\tau l}], \ \ \
\\ \nonumber
&&{\cal H}=Js^2\sum_{l'=<l>}{\bf n}_{a\tau l}\cdot{\bf n}_{b\tau l'}, \ \ \
{\bf n}_{a\tau l} \in a, \ \ \ {\bf n}_{b\tau l} \in b.
\end{eqnarray}
Now we
can introduce new more convenient variables
$\mbox{\boldmath{$\Omega$}}(\tau,l)$ and
${\bf M}(\tau,l)$ which realize the stereographic
mapping of a sphere:
\begin{eqnarray} \label{2.5}
{\bf n}_{a,b} = \frac{\pm \mbox{\boldmath{$\Omega $}}
\left(1-{\bf M}^2/4\right)-[\mbox{\boldmath{$\Omega $}}\times
{\bf M}]}{1+{\bf M}^2/4}, \ \ \ \mbox{\boldmath{$\Omega $}}^2=1, \ \ \
\mbox{\boldmath{$\Omega $}}\cdot{\bf M}=0.
\end{eqnarray}

In terms of these variables the total Lagrangian ${\cal L}_{\Omega M}=
{\cal L}_{kin}+{\cal H}$ has the final form
\begin{eqnarray} \label{2.5a}
&{\cal L}_{kin}=&\frac{2is\dot{\mbox{\boldmath{$\Omega$}}}\cdot{\bf M}}
{1+{\bf M}^2/4}, \ \ \ {\cal H}=
Js^2\sum_{l'=<l>}\{\mbox{\boldmath{$\Omega$}}\cdot
\mbox{\boldmath{$\Omega$}}'[(1-{\bf M}^2/4)(1-{\bf M}'^2/4)
\\ \nonumber
&&-{\bf M}\cdot{\bf M}']+
\mbox{\boldmath{$\Omega$}}\cdot{\bf M}' \
\mbox{\boldmath{$\Omega$}}'\cdot{\bf M}\}(1+{\bf M}^2/4)^{-1}
(1+{\bf M}'^2/4)^{-1},
\end{eqnarray}
where $\mbox{\boldmath{$\Omega$}}\equiv\mbox{\boldmath{$\Omega$}}
_{\tau l}, \ \mbox{\boldmath{$\Omega$}}'\equiv\mbox{\boldmath{$\Omega$}}
_{\tau l'}, \ {\bf M}\equiv{\bf M}_{\tau l}, \ {\bf M}'
\equiv{\bf M}_{\tau l'}$.
After this change of variables the measure of integration $D\mu({\bf n})$
(see (\ref{2.2b})) passes into
\begin{eqnarray} \label{2.6}
&&D\mu({\bf n})=\prod_{\tau l}\frac{(2s+1)^2}{4\pi^2}
\delta({\bf n}^2_{a\tau l}-1)\delta({\bf n}^2_{b\tau l}-1)
d{\bf n}_{a\tau l}d{\bf n}_{b\tau l} =
\\ \nonumber
&&=\prod_{\tau l}\frac{(2s+1)^2}{2\pi^2}
\frac{1-{\bf M}^2/4}{(1+{\bf M}^2/4)^3}
\delta\left(\mbox{\boldmath{$\Omega$}}^2-1\right)
\delta\left(\mbox{\boldmath{$\Omega$}}\cdot{\bf M}\right)
d\mbox{\boldmath{$\Omega$}}d{\bf M},
\end{eqnarray}
where the product in (\ref{2.6}) is performed over the AF
(doubled) lattice cells.

\subsection{Gauge transformation}
\label{sec:gauge}

The variable $\mbox{\boldmath{$\Omega$}}$ is responsible for the
AF fluctuations and the variable ${\bf M}$ for the
ferromagnetic ones. The ferromagnetic fluctuations are small
according to the
parameter $1/2s$ and therefore one can
expand the Lagrangian ${\cal L}_{\Omega M}$ (\ref{2.5a}) over ${\bf M}$.
The vector of the ferromagnetic fluctuations ${\bf M}$ plays the role (up to
factor 2s) of the canonical momentum conjugated to the canonical
coordinate $\mbox{\boldmath{$\Omega$}}$.  The term of first order in
${\bf M}$ coincides (after change of variables) with previous results
Manousakis \cite{Man1} and  Sachdev \cite{Sac2}.

 From Eq. (\ref{2.5a}) one can easily extract the quadratic part in
the variables $\mbox{\boldmath{$\Omega$}}$ and ${\bf M}$
of the total lagrangian, ${\cal L}_{quad}$,
\begin{eqnarray} \label{2.11}
{\cal L}_{quad} = 2is({\bf M}\cdot\dot{\mbox{\boldmath{$\Omega$}}}) +
Js^2\sum_{l'\in<l>}\left[
\mbox{\boldmath{$\Omega$}}^2-\mbox{\boldmath{$\Omega$}}\cdot
\mbox{\boldmath{$\Omega$}}' + {\bf M}^2 + {\bf M}\cdot{\bf M}'\right],
\end{eqnarray}

 The Lagrangian ${\cal L}_{quad}$ (\ref{2.11}) is
very simple but the measure $D\mu$ (\ref{2.6}) is not simple
due to the presence of two delta-- functions. Therefore we can not
simply perform the Gaussian integration over the field
$\mbox{\boldmath{$\Omega$}}$ and ${\bf M}$.

To solve this problem we shall use the method of the Lagrange multiplier
together with the saddle point approximation \cite{Pol1,Zin1} to eliminate
$\delta(\mbox{\boldmath{$\Omega$}}^2-1)$:
\begin{eqnarray} \label{2.11a}
&&\delta(\mbox{\boldmath{$\Omega$}}^2-1)=(2\pi)^{-1}
\int_{-\infty}^{\infty} \exp\left(A_{\lambda}\right)d\lambda,
\\ \nonumber
&&-A_{\lambda}={\cal L}_{\lambda}=(i\lambda+c_{\mu})
(\mbox{\boldmath{$\Omega$}}^2-1)
\end{eqnarray}
where $\lambda$ is the Lagrange multiplier, and  $c_{\mu}$ is a constant
which will be fixed with the help of the saddle point condition
\cite{Pol1,Zin1}.

To eliminate $\delta(\mbox{\boldmath{$\Omega$}}\cdot{\bf M})$ we shall
use some kind of the Faddev--Popov trick which was proposed in
\cite{Bel1}. Let us consider the integral:
\begin{eqnarray} \label{2.12}
I= \int f({\bf M})
\delta(\mbox{\boldmath{$\Omega $}}\cdot{\bf M})d{\bf M},
\end{eqnarray}
and insert it in the right hand side of the identity:
\begin{eqnarray} \label{2.13}
1=(\det(\hat{B}_{gaug}))^{1/2}\int \exp\left(-\frac{1}{2}
\varphi \hat{B}_{gaug} \varphi \right) d\varphi/\sqrt{2\pi},
\end{eqnarray}
where $\hat{B}_{gaug}$ is a positive number or
a positive definite operator for
some multi dimensional generalization.
After changing the order of integration
over ${\bf M}$ and $\varphi$
we can make the change of the variable ${\bf M}$:
${\bf M}\rightarrow {\bf M} - \mbox{\boldmath{$\Omega$}}\varphi$. After
that, due to the delta-function, we have
$\varphi = (\mbox{\boldmath{$\Omega$}}\cdot{\bf M})$ and the delta--function
disappears from the integral (\ref{2.12}):
\begin{eqnarray} \label{2.14}
&&I= (\det(\hat{B}_{gaug}))^{1/2}\int f({\bf M}_{tr})
\exp\left(A_{gaug}\right)d{\bf M}/\sqrt{2\pi}, \ \ \
\\ \nonumber
&&{\bf M}_{tr}\equiv{\bf M}-\mbox{\boldmath{$\Omega $}}
(\mbox{\boldmath{$\Omega $}}\cdot{\bf M}), \ \ \
-A_{gaug}={\cal L}_{gaug}=\frac{1}{2}(\mbox{\boldmath{$\Omega $}}
\cdot{\bf M})\hat{\hat{B}_{gaug}}(\mbox{\boldmath{$\Omega $}}\cdot{\bf M})
\end{eqnarray}
With the help of the identity (\ref{2.14}) we can remove the delta--function
$\delta (\mbox{\boldmath{$\Omega$}}\cdot{\bf M})$ from the measure (\ref{2.6}).
As a result we must substitute ${\bf M} \rightarrow {\bf M}_{tr}$ in the
Lagrangian ${\cal L}_{\Omega M}$ (\ref{2.5a}) and add the gauge fixing
Lagrangian ${\cal L}_{gaug}$ due to the additional exponent in (\ref{2.14}).
It is very convenient to chose the Lagrangian
${\cal L}_{gaug}$ in the form
\begin{eqnarray} \label{2.14a}
{\cal L}_{gaug}=Js^2\sum_{l'\in<l>}\left[
(\mbox{\boldmath{$\Omega$}}\cdot{\bf M})^2 +
(\mbox{\boldmath{$\Omega$}}\cdot{\bf M})
(\mbox{\boldmath{$\Omega$}}'\cdot{\bf M}')\right].
\end{eqnarray}
Such choice kills the major dependence on
$\mbox{\boldmath{$\Omega$}}$ in the Lagrangian (\ref{2.11}) which appears
due to substitution ${\bf M} \rightarrow {\bf M}_{tr}$.
We can also substitute ${\bf M}_{tr} \rightarrow {\bf M}$ in the first
term of the Lagrangian ${\cal L}_{quad}$ (\ref{2.11}) due to the identity
$(\mbox{\boldmath{$\Omega $}}\cdot\dot{\mbox{\boldmath{$\Omega$}}}) = 0$.
In this way, the expression (\ref{2.11}) for
${\cal L}_{quad}$ is valid in the
leading order with respect to $1/2s$.

The final expression for the GPF $Z$ of the Heisenberg AF is
\begin{eqnarray} \label{2.15}
&&Z = \int_{-\infty}^{\infty} \cdots
\int_{-\infty}^{\infty} D\mu(\mbox{\boldmath{$\Omega$}},{\bf M},\lambda)
\exp\left[A(\mbox{\boldmath{$\Omega $}},{\bf M},\lambda)\right],
\\ \nonumber
&&D\mu(\mbox{\boldmath{$\Omega$}},{\bf M},\lambda)=Z_{gaug}
D\mu(\mbox{\boldmath{$\Omega$}},{\bf M})D\mu(\lambda), \ \ \
Z_{gaug}=\left[\det(\hat{B}_{gaug})\right]^{1/2},
\\ \nonumber
&&D\mu(\mbox{\boldmath{$\Omega$}},{\bf M})=\prod_{\tau l}
\frac{(2s+1)^3(1-{\bf M}^2/4)}{(2\pi)^3(1+{\bf M}^2/4)^3}
d\mbox{\boldmath{$\Omega$}}(\tau,l)d{\bf M}(\tau,l),
\\ \nonumber
&&D\mu(\lambda)=\prod_{\tau l}\frac{2d\lambda(\tau,l)}{(2s+1)(2\pi)^{1/2}},
\end{eqnarray}
where $D\mu(\mbox{\boldmath{$\Omega$}},{\bf M},\lambda)$ is the measure
of an integration, and the action
$A(\mbox{\boldmath{$\Omega $}},{\bf M},\lambda)$ is determined by
the total Lagrangian ${\cal L}_{tot}= {\cal L}_{\Omega M} +
{\cal L}_{gaug}+{\cal L}_{\lambda}$ (\ref{2.5a},\ref{2.11a},\ref{2.14a}).
\subsection{Properties of the basic approximation}
\label{sec:d23}

 One can pass to the $q=(\omega,{\bf k})$ momentum representation
($\omega$ is the frequency, ${\bf k}$ is the wave vector),
and write out the total quadratic part of the Lagrangian in the matrix form
\begin{eqnarray} \label{2.15a}
&&{\cal L}_{quad}(q)=s{\bf X}^*_q
\hat{\Lambda}(q){\bf X}_q, \ \ \
{\bf X}^*_q=(\mbox{\boldmath{$\Omega$}}^*_q,{\bf M}^*_q),
\\ \nonumber
&&\hat{\Lambda}(q){\bf X}_q=
\left(\begin{array}{cc} P'_{\bf k},  & \omega  \\
-\omega, & Q_{\bf k} \end{array} \right)
\left(\begin{array}{c} \mbox{\boldmath{$\Omega$}}_q \\
{\bf M}_q\end{array} \right), \ \ \ P'_{\bf k}=P_{\bf k}+c_{\mu}, \ \ \
c_{\mu}=\mu_0^2/2{\cal J},
\\ \nonumber
&&(Q_{\bf k},P_{\bf k}) = {\cal J}(1 \pm \gamma_{\bf k}), \ \ \
\gamma_{\bf k} = (1/2)(\cos(k_xa)+\cos(k_ya)),\ \ \ {\cal J}=Jsz.
\end{eqnarray}
Here ${\bf X}_q$ is a two component vector field which combines the vector
fields $\mbox{\boldmath{$\Omega$}}_q$ and ${\bf M}_q$; the constant
$c_{\mu}$ (\ref{2.11a}) is expressed through the constant $\mu_0$ which is
the mass of the $\mbox{\boldmath{$\Omega$}}$ field  in the lowest order of perturbation
theory. One can invert the $2\times 2$ matrix $\hat{\Lambda}(q)$ and
get the bare Green function $\hat{G}_q$ of the
$\mbox{\boldmath{$\Omega$}}_q$ and ${\bf M}_q$ fields
\begin{eqnarray} \label{2.15b}
&&\hat{G}_q\equiv \left(\begin{array}{cc} G^{\Omega}_q, & G^d_q \\
G_q^u, & G^M_q \end{array} \right)=
\frac{1}{2s}\left(\hat{\Lambda}(q)\right)^{-1}=
\frac{1}{2sL_q} \left(\begin{array}{cc} Q_{\bf k}, & -\omega \\
\omega, & P'_{\bf k}\end{array} \right),
\\ \nonumber
&&L_q=\omega^2+\omega^2_{0{\bf k}}, \ \ \
\omega^2_{0{\bf k}}=P'_{\bf k}Q_{\bf k}=
(1-\gamma^2_{\bf k}){\cal J}^2+(1+\gamma_{\bf k})\mu_0^2/2.
\end{eqnarray}
Here $\omega_{0{\bf k}}$ is the primary magnon frequency in the
paramagnetic phase, $G^{\Omega}_q, \  G^d_q, \ G_q^u,\ G^M_q$ are
notations for the matrix elements of the matrix Green function
$\hat{G}_q$.

At first let us discuss the parameter of the perturbation theory.
One can see from an explicit form of the Lagrangian (\ref{2.5a}) that the
spin wave nonlinearity of the theory is caused by the term ${\bf M}_{tr}$
and its modifications. Its average value is
\begin{eqnarray} \label{2.16}
<{\bf M}^2_{tr}> = (N-1)\sum_{q}G^M_q =\frac{(N-1)T}{2s}
\sum_{\omega=2\pi nT,{\bf k}}\frac{P_{\bf k}}
{\omega^2+\omega^2_{0{\bf k}}},
\end{eqnarray}
where $N=3$ is the number of components of the $\mbox{\boldmath{$\Omega$}}$ field.
Summation over $\omega$ is
obtained by standard methods \cite{Zin1} and we have
\begin{eqnarray} \label{2.17}
<{\bf M}^2_{tr}>=\frac{(N-1)}{2s}\sum_{\bf k}
\frac{P_{\bf k}}{2\omega_{0{\bf k}}}(1+2n_{0{\bf k}})=
\frac{(N-1)}{4s}\left\{\begin{array}{cc}
C_{M0}, & T \ll {\cal J} \\
(T/{\cal J})C_{M\infty}, & T \gg {\cal J},
\end{array} \right..
\end{eqnarray}
Here, $N=3$ is the number of components of fields ${\bf M}$ and
$\mbox{\boldmath{$\Omega $}}$, $n_{0{\bf k}}=\left(\exp(\omega_{0{\bf
k}}/T)-1\right)^{-1}$ is the Plank function, and
summation over ${\bf k}$
means the normalized integration over the AF Brillouin band:
\begin{eqnarray} \label{2.18}
\sum_{\bf k} f({\bf k})\equiv \int\frac{a^2d{\bf k}}{2\pi^2} f({\bf k}).
\end{eqnarray}
The constants $C_{M0}$ and $C_{M\infty}$ are defined by the relations
\begin{eqnarray} \label{2.18a}
&&C_{M0}=\sum_{\bf k}\sqrt{\frac{1-\gamma_{\bf k}}{1+\gamma_{\bf k}}}
=c_0-c_1=0.65075,\ \ \
C_{M\infty}=\sum_{\bf k}\frac{2}{1+\gamma_{\bf k}}=1.48491,
\end{eqnarray}
where all sum of the type of (\ref{2.18}) are calculated by the following
method
\begin{eqnarray} \label{2.19}
&&\sum_{\bf k}f(\gamma_{\bf k})=
\int_{0}^{1}f(\epsilon) \rho(\epsilon)d\epsilon, \ \ \
c_n=\sum_{\bf k}\frac{\gamma^n_{\bf k}}{\sqrt{1-\gamma^2_{\bf k}}},
\\ \nonumber
&&\rho (\epsilon)=\sum_{\bf k}
\delta(\epsilon-\gamma_{\bf k})=\frac{4}{\pi^2}K(1-\epsilon^2),
\end{eqnarray}
where $K(x)$ is the complete elliptic integral of the first kind.
By the same manner one can calculate the different time and space
average
\begin{eqnarray} \label{2.20}
&&<M_i(\tau+\delta,{\bf l}+{\bf r})\Omega_j(\tau,{\bf l})> =
\delta_{ij}\frac{T}{2s}
\sum_{\omega=2\pi nT,{\bf k}}e^{i\delta\tau +
i{\bf k}\cdot{\bf r}}\frac{\omega} {\omega^2+\omega^2_{0{\bf k}}}=
\\ \nonumber
&&=\frac{i}{4s}\delta_{ij}\epsilon(\delta)\sum_{\bf k}
\frac{\exp(i{\bf k}\cdot{\bf r}-|\delta|\omega_{0{\bf k}})}
{1-\exp(-\beta\omega_{0{\bf k}})}.
\end{eqnarray}
Here $\epsilon(\delta)$ is the sign function: $\epsilon(\delta)=1$ for
$\delta >0$, $\epsilon(\delta)=-1$ for $\delta <0$. For ${\bf r}=0$ and
$\delta{\cal J} \ll 1$ we have an explicit expression
\begin{eqnarray} \label{2.20a}
<M_i(\tau+\delta,{\bf l})\Omega_j(\tau,{\bf l})> =
\frac{i}{4s}\delta_{ij}\epsilon(\delta)
\left\{\begin{array}{cc}
1,& T \ll {\cal J} \\
(T/{\cal J})c_0, & T \gg {\cal J},
\end{array} \right..
\end{eqnarray}

 From Eqs. (\ref{2.17},\ref{2.20}) we clearly see that the
 parameter of perturbation theory is $1/2s$ at low
temperatures $T\ll{\cal J}$ and $T/(2s{\cal J})$ at high temperatures
$T\ge{\cal J}$.
Thus, perturbation theory is working when $1/2s$ is a
small parameter and the temperature is not high. We remind
the reader that this is just the applicability condition of the spin wave
theory.

 Now we can consider the saddle point condition for the $\lambda$ field
$<\mbox{\boldmath{$\Omega$}}^2>=1$ which is the
most important constraint of the theory which determines its phase state:
\begin{eqnarray} \label{2.21}
1=<\mbox{\boldmath{$\Omega$}}^2>=N\sum_{q}G^{\Omega}_q =
\frac{NT}{2s}\sum_{\omega=2\pi nT}
\sum_{\bf k}\frac{Q_{\bf k}}{\omega^2+\omega^2_{0{\bf k}}},
\end{eqnarray}
\begin{eqnarray} \label{2.22}
1=<\mbox{\boldmath{$\Omega$}}^2>=\frac{N}{2s}\sum_{\bf k}
\frac{Q_{\bf k}}{2\omega_{0{\bf k}}}(1+2n_{0{\bf k}}),
\end{eqnarray}
 The right hand side of Eq. (\ref{2.22}) contains
two terms. The first term $Q_{\bf k}/2\omega_{0{\bf k}}$ is responsible
for the quantum fluctuations of the $\mbox{\boldmath{$\Omega$}}$ field.The second term
$Q_{\bf k}n_{0{\bf k}}/\omega_{0{\bf k}}$ is responsible for the classical
thermal fluctuations of the $\mbox{\boldmath{$\Omega$}}$ field. The role of these two terms
is quite different. The quantum fluctuations are small with respect to the
parameter of perturbation theory $1/2s$ and for basic approximation
they can be neglected. The quantum fluctuation can be considered in the
continuum approximation for which $Q_{\bf k}\simeq 2{\cal J}$ and
$\omega^2_{0{\bf k}}\simeq c^2_{0s}{\bf k}^2+\mu^2$, where $c_{0s}$ is the
primary velocity of sound and $c^2_{0s}= 2J^2s^2za^2/\hbar^2$.
The integration over the two dimensional momentum ${\bf k}$
can be easily performed  \cite{Sac1}). The integration over the angle
is trivial, and integration over the modulus $|{\bf k}|$ is performed if
we introduce new variable of integration $x=\beta\omega_{0{\bf k}}$.
As a result we have the constraint condition
\begin{eqnarray} \label{2.23}
1=-\frac{TN}{2\pi Js^2}\ln(1-\exp(-\mu/T))\simeq
-\frac{TN}{2\pi Js^2}\ln(\mu/T)).
\end{eqnarray}
The coefficient before the logarithm in this equation is always small when the
regime of the weak coupling is valid. At small temperatures $T\ll {\cal J}$
this is obvious. At the temperature $T\simeq {\cal J}$ this coefficient
coincides with the parameter of perturbation theory (\ref{2.17}) and also
must be considered as small. This means the logarithm in (\ref{2.23}) must
be negative and large in modulus. This leads to the condition
$\mu/T \ll 1$ which justifies the last simplification in (\ref{2.23}).
As a result we have the well known \cite{Man1,Cha1,Sac1} zero order
expression for $\mu$
\begin{eqnarray} \label{2.24}
\mu = T\exp\left(-\frac{2\pi Js^2}{TN}\right), \ \ \
\xi = \hbar c_{s}/\mu,
\end{eqnarray}
where $\xi$ is the correlation length. From Eq. (\ref{2.24}), the very
important conclusion  follows: {\it in the regime of the weak
coupling the correlation length} $\xi$ {\it is much larger than the
lattice constant} $a$. This conclusion makes possible the scale
separation for the problem of disordered QAF
\cite{Cha1}.

 To close the theory it is helpful to define the polarization operator
$\Pi(q)$ of the $\mbox{\boldmath{$\Omega$}}$ field
\begin{eqnarray} \label{2.25}
A_{\lambda quad}=-\frac{1}{2}\sum_{q}\lambda^*(q)\Pi(q)\lambda(q),
\end{eqnarray}
which is simply a loop, and the Green function of the $\lambda$
field is $\Pi(q)^{-1}$. In the lowest approximation $\Pi(q)$ is simply
a loop from two Green function $G^{\Omega}$
\begin{eqnarray} \label{2.26}
\Pi_0(q)=2NT\sum_{q'}G^{\Omega}(q')G^{\Omega}(q-q').
\end{eqnarray}
Using the Green function $G^{\Omega}(q)$ from (\ref{2.15b}) we can perform the
summation over $\omega$  and have  the expression for the simple loop
\begin{eqnarray} \label{2.27}
\Pi_0(q)=\frac{N}{4s^2}\sum_{\bf k'}(1+2n_1)\frac{Q_1Q_2}{\omega_1\omega_2}
\left[\frac{\omega_2+\omega_1}{\omega^2+(\omega_2+\omega_1)^2}+
\frac{\omega_2-\omega_1}{\omega^2+(\omega_2-\omega_1)^2}\right],
\end{eqnarray}
where the index $i:=1,2$ corresponds the momentum ${\bf k}_1\equiv{\bf k}'$
and ${\bf k}_2\equiv{\bf k}-{\bf k}'$. The main contribution in $1/2s$
in (\ref{2.27}) is from the thermal fluctuations even at low temperatures
$T$, because the integral strength of such fluctuations is fixed by the
saddle point condition (\ref{2.21}) and does not depend on temperature.
The explicit form for $\Pi_0(q)$
may be obtained in two limiting cases $\hbar q \gg T$ and $\hbar q \ll T$,
where $q^2=\omega^2+c^2_s k^2$. In the first case the momentum
$k'\sim T/c_s\ll q$, and we can separate the integration over
${\bf k}'$ and put ${\bf k}'=0$ in all 
places in (\ref{2.27}) affected by the factor 
$n_1\equiv n({\bf k}')$. The result is extremely simple
\begin{eqnarray} \label{2.28}
\Pi_0(q)=4G^{\Omega}(q)=\frac{2{\cal J}(1+\gamma_{\bf k})}
{s(\omega^2+\omega^2_{0{\bf k}})}, \ \ \ q \gg k_T,\ \ \ k_T=T/c_s.
\end{eqnarray}
At small $q \ll c_s/a$ a similar result was obtained in \cite{Sac1}.
Notice, that it exceeds the quantum contribution in (\ref{2.27})
$\Pi_0(q)=N/4q$ in the large parameter $16s{\cal J}/Nq$. The second limiting
case $\hbar q \ll T$ lies in the purely continuum region. It corresponds
to pure classical two dimensional case: $\omega=0$, and $\omega'=0$.
Integration over ${\bf k}'$ can be easily performed, the result
coincides with \cite{Sac1} up to the normalization factor:
\begin{eqnarray} \label{2.28a}
\Pi_0(k)=\frac{8NT}{\pi s^2q\sqrt{q^2+4\mu^2}}
\ln\left(\frac{q+\sqrt{q^2+4\mu^2}}{2\mu}\right), \ \ \ q=c_s k.
\end{eqnarray}
\subsection{The spin correlation functions}
\label{sec:spic}

 The approach of this paper 
allows us to find the spin
correlation functions in all values of $\omega$ and ${\bf k}$.
The dynamical spin susceptibility is determined by the relation
\begin{eqnarray} \label{2.29}
&&\chi(\omega,{\bf k})\delta_{ij} = -\frac{i}{\hbar}\int_{0}^{\infty}
d\tau \sum_{\bf l} Tr\left\{\left[\hat{S}_i(\tau,{\bf l}),
\hat{S}_j(0,{\bf 0})\right]\right.\cdot
\\ \nonumber
&&\cdot\left.\exp\left(\beta(F-\hat{H})\right)\right\}
\exp\left(i\omega\tau
-ia{\bf k}\cdot{\bf l}\right),
\end{eqnarray}
where $F=-T\ln(Z)$ is the free energy, and the wave vector ${\bf k}$ runs
without limitations over the Brillouin band.  It is well known that
the dynamical spin susceptibility $\chi(\omega,{\bf k})$ coincides with the
temperature Green function continued on the imaginary frequency $\omega$.
It can be calculated on the basis of the functional integral (\ref{2.15})
\begin{eqnarray} \label{2.30}
\chi(\omega,{\bf k})\delta_{ij}\delta({\omega-\omega'})
\delta({{\bf k}-{\bf k}'})= -\frac{is^2}{\hbar}
<n^*_i(\mbox{\boldmath{$\Omega$}},{\bf M},\omega',{\bf k}')
n_j(\mbox{\boldmath{$\Omega$}},{\bf M},\omega,{\bf k})>,
\end{eqnarray}

Here, the unit vector ${\bf n}(\mbox{\boldmath{$\Omega$}},{\bf M},
\omega,{\bf k})$ is a function of the fields
$\mbox{\boldmath{$\Omega$}}(\tau,l),{\bf M}(\tau,l)$ according
to (\ref{2.5}); the  brackets $<...>$ mean averaging over the
$\mbox{\boldmath{$\Omega$}}(\tau,l),{\bf M}(\tau,l)$ fields according
to (\ref{2.15}).
Eq. (\ref{2.30}) reduces the problem of calculation of
the spin Green function to the problem of the calculation of the averages
of the $\mbox{\boldmath{$\Omega$}}$
and ${\bf M}$ fields. In the lowest order in $1/2s$ it is sufficient
to  use the lowest order relation
\begin{equation}\label{2.31}
{\bf n}(\mbox{\boldmath{$\Omega$}}(\tau,l),{\bf M}(\tau,l),\tau,l)
\simeq e^{ia{\bf l}\cdot{\bf q}_{AF}}
\mbox{\boldmath{$\Omega$}}(\tau,l)-
[\mbox{\boldmath{$\Omega $}}(\tau,l)\times {\bf M}(\tau,l)],
\end{equation}
where ${\bf q}_{AF}=(\pi/a,\pi/a)$ is the AF vector (\ref{2.5}).
Substituting  the vector ${\bf n}$ from (\ref{2.31}) into (\ref{2.30}) we get
the dynamical spin susceptibility as a sum of two terms
$\chi(\omega,{\bf k})=\chi_A(\omega,{\bf k})+\chi_F(\omega,{\bf k})$.
The spin susceptibility $\chi_A(\omega,{\bf k})$ is responsible for the
AF fluctuations. It is proportional to the Green function
$G^{\Omega}_{q}$ analytically
continued and shifted by the AF vector ${\bf q}_{AF}$
\begin{eqnarray} \label{2.32}
\chi_A(\omega,{\bf k})=-\frac{Js^2z(1+\gamma_{{\bf k}^*})}
{2(\omega^2-\omega^2_{0{\bf k}^*}+i\omega\delta)},
\end{eqnarray}
where ${\bf k}^*={\bf k}-{\bf q}_{AF}$, $\omega^2_{0{\bf k}}$ is the
magnon frequency (\ref{2.15b}).
For the spin susceptibility $\chi_A(\omega,{\bf k})$ we have a loop
expression
\begin{eqnarray} \label{2.33}
\chi_F(\omega,{\bf k})=2s^2 \sum_{q'}\left[G^{\Omega}(q')G^{M}(q-q')-
G^{u}(q')G^{d}(q-q')\right].
\end{eqnarray}
In the main approximation in $1/2s$ the first term gives the main
contribution in the same manner as in (\ref{2.26}) when the contribution from
the Plank function $n_1$ was dominated \begin{eqnarray} \label{2.34}
\chi_F(\omega,{\bf k}) \simeq -\sum_{\bf k'}n_1
\frac{Q_1P'_2}{2\omega_1\omega_2}
\left[\frac{\omega_2+\omega_1}{\omega^2-(\omega_2+\omega_1)^2+i\omega\delta}+
\frac{\omega_2-\omega_1}{\omega^2-(\omega_2-\omega_1)^2+i\omega\delta}\right],
\end{eqnarray}
where the notation is the same as in (\ref{2.26}). In the case of
$q=\sqrt{\omega^2-c^2_s k^2} \gg k_T$ the expression (\ref{2.34}) is
substantially simplified on the basis of the idea of dominant small
$k'\simeq k_T$,
\begin{eqnarray} \label{2.35}
\chi_F(\omega,{\bf k}) \simeq  -\frac{2s^2}{N}G^M(q)=
-\frac{Js^2z(1-\gamma_{\bf k})}
{N(\omega^2-\omega^2_{0{\bf k}}+i\omega\delta)}.
\end{eqnarray}
For the case $q \le k_T$ the expression for $\chi_F(\omega,{\bf k})$ is
not so simple and we present the result the limit
$\mu \ll \omega,\ \omega_{\bf k} \ll k_T$
\begin{eqnarray} \label{2.36}
&&\chi_F(\omega,{\bf k})=
-\frac{a^2T^2}{\pi c_s^2q}\xi(2)+ i\epsilon(\omega)\frac{a^2T}
{16\pi c_s^2|q|^2}\cdot
\\ \nonumber
&&\cdot\left(2\pi\theta(q^2)\omega(4\omega-3|q|)+
\theta(-q^2)T|q|\xi(2)\right),
\end{eqnarray}
where $\xi(n)$ is the Riemann zeta function, and $\epsilon(\omega)$ is the
sign function. The two terms in (\ref{2.36}) which are proportional to
$\xi(2)$ are generated by the integral which is cut at large $k$ by the
Plank distribution function $n({\bf k}')$. The ferromagnetic spin
susceptibility (\ref{2.37}) is suppressed in comparison with the
antiferromagnetic one (\ref{2.32}) by the parameter $(qa)^2$.
\subsection{The spin-rotator model}
\label{sec:sprot}

 In the basic approximation in $1/2s$ the Lagrangian is quadratic in the
${\bf M}$ field and one can integrate over the field ${\bf M}$
and obtain the final action
for the field $\mbox{\boldmath{$\Omega$}}$ and the Lagrange multiplier
$\lambda$:
\begin{eqnarray} \label{2.37}
{\cal L}_{\Omega\lambda} = s(\dot{\mbox{\boldmath{$\Omega$}}}
\hat{Q}^{-1}\dot{\mbox{\boldmath{$\Omega$}}}) +
s(\mbox{\boldmath{$\Omega$}}\hat{P}'\mbox{\boldmath{$\Omega$}}) +
+i\lambda\left(\mbox{\boldmath{$\Omega$}}^2-1\right),
\end{eqnarray}
where the quantities $\hat{Q}$ and $\hat{P}'$ are defined in the ${\bf k}$
representation in (\ref{2.15b}). As a result of such integration
the ${\bf M}$ field is a function of the $\mbox{\boldmath{$\Omega$}}$ field:
${\bf M}= \hat{Q}^{-1}\dot{\mbox{\boldmath{$\Omega$}}}$.
 One can easily recognize in (\ref{2.16}) the Lagrangian of some kind of
the Quantum Lattice Rotator Model (QLRM)\cite{Cha1}. However it is different
from the standard model due to the momentum dependence of the kinetic term
in Eq. (\ref{2.16}). We shall call such kind of models spin--rotator
(SR) models.  The SR model describes a quantum antiferromagnet in the limit
$s\rightarrow\infty$. The QLRM is also well defined and does not contain
any divergences. It allows to perform all calculations accurately
because all physical quantities are well defined in the framework of this
model.

\section{Beyond the continuum approximation in time }
\label{sec:d3}
\subsection{Basic approach}
\label{sec:bas3}

When we try to construct the perturbation corrections to the basic
approach discussed in the previous section we meet a fundamental
difficulty: the integrals arising from the Green functions (\ref{2.15b}), over the
frequency $\omega$, are not well defined. This is obvious from
the consideration of the average $<M_i(\tau+\delta,{\bf l})
\Omega_j(\tau,{\bf l})>$. The result essentially depends on the time shift
$\delta$ (\ref{2.20}) which reflects the phase space nature of the
$\mbox{\boldmath{$\Omega$}}$ and ${\bf M}$ variables: the Green functions
$G^{u,d}(q) \sim \omega^{-1}$ at large $\omega$. Moreover there are some
doubts that the quantity $<{\bf M}^2_{tr}>$ was calculated completely
correctly. Actually, at low temperature \begin{eqnarray} \label{3.1} <{\bf
M}^2_{tr}> \sim \sum_{\bf k}\int_{-\pi/\Delta}^{\pi/\Delta}
\frac{d\omega}{2\pi}
\frac{{\cal J}(1-\gamma_{\bf k})}{\omega^2+\omega^2_{0{\bf k}}}=C_{M0},
\end{eqnarray}
where $\Delta$ is the time step in the accurate definition of the GPF
presented in the Appendix A (\ref{A1}), the interval
$(-\pi/\Delta,\pi/\Delta)$ gives the one dimensional Brillouin band for
the final time step $\Delta$. The limit $\Delta \rightarrow 0$ gives a
correct value of this average.  At first sight this limit is trivial
because the integral over $\omega$ in (\ref{3.1}) is well defined.  Suppose
that we were not so accurate and the numerator of (\ref{3.1}) in fact
contains a small corrections ${\cal J}(1-\gamma_{\bf k}) \Rightarrow {\cal
J}(1-\gamma_{\bf k}) +\Delta\omega^2$. If now we at first calculate the
integral (\ref{3.1}) with this numerator and after that pass to the limit
$\Delta \rightarrow 0$ the result will be different:
$C_{M0} \rightarrow C_{M0}+1$.
This example should convinced the reader that for a spin system the
continuum limit of the path integral for the GPS (\ref{2.2}-\ref{2.2b}) must
be produced with sufficient accuracy: at first it is necessary to formulate
the theory at finite $\Delta$ and one  can put $\Delta =0$ only after the
calculation of all integrals over $\omega$ has been done.
Notice also that the accurate
version, discrete in time, of the GPS (\ref{2.2}-\ref{2.2b}) is necessary
when we calculate the free energy.

 Now, on the basis of the results of the Appendix A we get an accurate expression
for the quadratic part of the action. Instead of the expression (\ref{2.3})
for the action $A({\bf n})$ we shall use a more accurate expression
\begin{eqnarray} \label{3.2}
A({\bf n}) = -\sum_{j=0}^{N_{\tau}}\sum_{l}\Delta\left[{\cal L}_{kin}(j,l) +
{\cal H}(j,l)\right],
\end{eqnarray}
where $\tau=j\Delta$, and $\Delta N_{\tau}=\beta$. According to the results of
the Appendix A, ${\cal L}_{kin}(j,l)$ consist of two parts
${\cal L}_{kin}={\cal L}_{mod}+{\cal L}_{pha}$. The
first term is pure real the second term is pure imaginary.

According to Eq. (\ref{A15}) the Lagrangian ${\cal L}_{mod}$ can presented
for two sublattices ${\rm a}$ and ${\rm b}$ in the form
\begin{eqnarray} \label{3.3}
\Delta{\cal L}_{mod}=-s\ln\left[(1+\underline{\bf n}_a\cdot
{\bf n}_a)(1+\underline{\bf n}_b\cdot{\bf n}_b)/
4\right],
\end{eqnarray}
where ${\bf n}_a={\bf n}_a(j,l)$, $\underline{\bf n}_a={\bf n}_a(j+1,l)$,
$\underline{\bf n}_b={\bf n}_b(j+1,l)$, ${\bf n}_b={\bf n}_b(j,l)$.
It is assumed that vectors $\underline{\bf n}_a, \ {\bf n}_a, \
\underline{\bf n}_b, \ {\bf n}_b$ are  functions of the dynamical
variables $\mbox{\boldmath{$\Omega $}}$ and ${\bf M}$
according to Eq. (\ref{2.5}).

The Lagrangian ${\cal L}_{pha}$ is not so simple and according to
Appendix A (\ref{A26}) it is
\begin{eqnarray} \label{3.4}
\Delta{\cal L}_{pha}=-\frac{s}{2}\ln\left(\frac{T_a\underline{T}_a^*
T_b\underline{T}_b^*}
{T_a^*\underline{T}_aT_b^*\underline{T}_b}\right),
\end{eqnarray}
where the quantity $T_{a,b}$ is defined in (\ref{A28}) and has a rather
complicated form. The 
expansion of $\Delta{\cal L}_{pha}$
over the field ${\bf M}$ contains only odd powers of ${\bf M}$.

The Hamiltonian ${\cal H}({\bf n})$ can be
obtained on the basis of  Eq.(\ref{A12}) for the matrix element of the spin
operator ${\bf S}$ if we substitute them in the Heisenberg Hamiltonian
\begin{eqnarray} \label{3.5}
{\cal H}({\bf n})=Js^2\sum_{l'\in <l>}
\mbox{\boldmath{${\cal  S}$}}(\underline{\bf n},{\bf n})
\cdot\mbox{\boldmath{${\cal  S}$}}
(\underline{\bf n}',{\bf n}'),
\end{eqnarray}
where $\underline{\bf n}={\bf n}_a(j+1,l)$, ${\bf n}={\bf n}_a(j,l)$,
$\underline{\bf n}'={\bf n}_b(j+1,l')$, ${\bf n}'={\bf n}_b(j,l')$.
All these vectors are also the functions of the dynamical
variables $\mbox{\boldmath{$\Omega $}}$ and ${\bf M}$
according to Eq. (\ref{2.5}).
The vector $\mbox{\boldmath{${\cal  S}$}}(\underline{\bf n},{\bf n})$
is determined by the relation
\begin{eqnarray} \label{3.6}
\mbox{\boldmath{${\cal  S}$}}(\underline{\bf n},{\bf n})=
\frac{<\underline{\bf n}|\hat{\bf S}|{\bf n}>}{<\underline{\bf n}
| {\bf n}>} = \frac{\underline{\bf n}+{\bf n}
-i[\underline{\bf n}\times{\bf n}]}{1+\underline{\bf n}\cdot{\bf n}},
\end{eqnarray}
and it transforms as a vector under rotations according to the discussion
in Appendix A.

Expanding the Lagrangians ${\cal L}_{mod}$ (\ref{3.3}), ${\cal L}_{pha}$
(\ref{3.4}), and the Hamiltonian (\ref{3.5}) in powers of the vector
${\bf M}$ up to second order we get
\begin{eqnarray} \label{3.7}
&&\Delta{\cal L}_{kin}=s[1-\underline{\mbox{\boldmath{$\Omega$}}}\cdot
\mbox{\boldmath{$\Omega$}}+{\bf M}^2-\underline{\bf M}\cdot{\bf M}+
i(\underline{\mbox{\boldmath{$\Omega$}}}\cdot{\bf M}-
\mbox{\boldmath{$\Omega$}}\cdot\underline{\bf M})], \ \ \
\\ \nonumber
&&{\cal H}=Js^2\sum_{l'\in <l>}[\underline{\mbox{\boldmath{$\Omega$}}}\cdot
\mbox{\boldmath{$\Omega$}}-\mbox{\boldmath{$\Omega$}}\cdot
\mbox{\boldmath{$\Omega$}}'+\underline{\bf M}\cdot{\bf M}+
{\bf M}\cdot{\bf M}'-i(\underline{\mbox{\boldmath{$\Omega$}}}\cdot{\bf M}-
\mbox{\boldmath{$\Omega$}}\cdot\underline{\bf M})],
\end{eqnarray}
where the usual quantities $\mbox{\boldmath{$\Omega$}}$ and ${\bf M}$
are for the arguments $j,l$; the underlined ones are for $j+1,l$;
with prime are for $j,l'$; the underlined with prime are for $j+1,l'$.

According to the analysis performed in the section (\ref{sec:gauge})
it is necessary to add to the Lagrangian (\ref{3.7}) the gauge Lagrangian
${\cal L}_{gaug}$ generalizing (\ref{2.14a}) in the case of finite time step
\begin{eqnarray} \label{3.8}
&&\Delta{\cal L}_{gaug}=s\left[(\mbox{\boldmath{$\Omega$}}\cdot{\bf M})^2-
(\underline{\mbox{\boldmath{$\Omega$}}}\cdot\underline{\bf M})
(\mbox{\boldmath{$\Omega$}}\cdot{\bf M})\right]+
\\ \nonumber
&&+\Delta Js^2\sum_{l'\in<l>}\left[
(\underline{\mbox{\boldmath{$\Omega$}}}\cdot\underline{\bf M})
(\mbox{\boldmath{$\Omega$}}\cdot{\bf M}) +
(\mbox{\boldmath{$\Omega$}}\cdot{\bf M})
(\mbox{\boldmath{$\Omega$}}'\cdot{\bf M}')\right].
\end{eqnarray}
This Lagrangian ${\cal L}_{gaug}$ kills the most strong interaction
between the $\mbox{\boldmath{$\Omega$}}$ and ${\bf M}$ fields.

In this step we can pass to the $q=(\omega, {\bf k})$ representation
\begin{eqnarray} \label{3.9}
&&{\bf X}(q))=
\sqrt{2/N_t}
\sum_{j,l} e^{ia{\bf k}\cdot{\bf l}-i\Delta\omega j}{\bf X}(j,l), \ \ \
N_t=N_{\tau}N_s
\\ \nonumber
&&\omega:=2\pi Tj, \ (-N_{\tau}+1)/2 <j<(N_{\tau}-1)/2, \ \ \
\\ \nonumber
&&{\bf k}=\left(\frac{2\pi m_a}{bN_a}, \frac{2\pi m_b}{bN_b}\right), \ \ \
-\frac{N_{a,b}-1}{2} <m_{a,b}<\frac{N_{a,b}-1}{2},
\end{eqnarray}
where we choose $N_{\tau},N_a,N_b$ as odd numbers, $N_s=N_aN_b$ is
the number of sites on one sublattice, $j,m_a,m_b$ are natural numbers.

 Now one can write the action in the form $A(\mbox{\boldmath{$\Omega$}},
{\bf M})=\sum_{q}{\cal L}_{quad}(q)$ where the Lagrangian
${\cal L}_{quad}(q)$ is presented in (\ref{2.15a}) but the matrix
$\hat{\Lambda}(q)$ acting on ${\bf X}_q=(\mbox{\boldmath{$\Omega$}}_q,
{\bf M}_q)$ is now different
\begin{eqnarray} \label{3.11}
&&\hat{\Lambda}(q) = \left(\begin{array}{cc} \Lambda^{\Omega}(q), &
\Lambda^{u}(q)\\
-\Lambda^{u}(q), & \Lambda^{M}(q)\end{array}\right), \ \ \
\Lambda^{M}(q)=1-c_{\omega}+ \Delta{\cal J}(c_{\omega}+\gamma_{\bf k}), \ \ \
\\ \nonumber
&&\Lambda^{\Omega}(q)=1-c_{\omega}+\Delta{\cal J}(c_{\omega}-
\gamma_{\bf k})+\Delta\mu_0^2/2{\cal J}, \ \ \
\Lambda^{u}(q)=s_{\omega}(1-\Delta{\cal J})
\end{eqnarray}
where $c_{\omega}=\cos(\omega\Delta)$ and $s_{\omega}=\sin(\omega\Delta)$.
Inverting the matrix $\hat{\Lambda}(q)$ we get the Green function
generalizing (\ref{2.15b}) in the case of finite time step $\Delta$
\begin{eqnarray} \label{3.12}
&&\hat{G}_q=\left(\hat{\Lambda}(q)\right)^{-1}=
\frac{1}{2s\bar{L}(q)}
\left(\begin{array}{cc} \Lambda^{M}(q), & -\Lambda^{u}(q) \\
\Lambda^{u}(q),& \Lambda^{\Omega}(q)\end{array} \right),
\\ \nonumber
&&\bar{L}(q)\simeq(1-\Delta {\cal J}+\Delta\mu_0^2/4{\cal J})
[2(1-c_{\omega})+\Delta^2 \omega^2_{0{\bf k}}],
\end{eqnarray}
where the quantity $Q_{\bf k}$, $P'_{\bf k}$, and the bare frequency
$\omega_{0{\bf k}}$ were defined in (\ref{2.15b}). At small $\omega$ when
$\omega\Delta \ll 1$ if we neglect small terms of order $\Delta{\cal J}$ and
$\Delta\mu^2_0/{\cal J}$ the matrices $\hat{\Lambda}(q)$ and $\hat{G}_q$
pass (up to normalization factor $\Delta$ and $\Delta^{-1}$) into their
continuum analogues (\ref{2.15a}) and (\ref{2.15b}).

Of course, the difference between the continuum expressions
(\ref{2.15a}) and (\ref{2.15b}) for $\hat{\Lambda}(q)$ and $\hat{G}_q$
and the precise values (\ref{3.11}) and (\ref{3.12}) are only essential for the
intermediate steps of the calculations. We want to stress that the
amplitudes of the magnon scattering in the skeleton approximation are
determined by the continuum action (\ref{2.15}) and the continuum Green
function (\ref{2.15b}) only.

 At first we demonstrate that the simple averages calculated in the section
(\ref{sec:d23}) in fact were not calculated with sufficient accuracy.
The tricks necessary to carry out the calculation are discussed in the
Appendices. The result at low $T\ll {\cal J}$ are presented below.
\begin{eqnarray} \label{3.13}
& &<M_iM_j>= \frac{1}{4s}\delta_{ij}(1+c_0-c_1), \hskip 1.cm
<\underline{M}_iM_j>=\frac{1}{4s}\delta_{ij}(c_0-c_1),
\\ \nonumber
&&<M'_iM_j>=\frac{1}{4s}\delta_{ij}(\frac{4}{\pi^2}+c_1-c_2), \hskip 1.cm
<\underline{M}_iM'_j>=\frac{1}{4s}\delta_{ij}(c_1-c_2),
\\ \nonumber
&&<\Omega_i\underline{M}_j>=i\frac{1}{4s}\delta_{ij}, \hskip 3.cm
<\underline{\Omega}_iM_j>=-i\frac{1}{4s}\delta_{ij}
\\ \nonumber
&&<\Omega'_i\underline{M}_j>=i\frac{1}{\pi^2s}\delta_{ij}, \hskip 3.cm
<\underline{\Omega}_iM'_j>=-i\frac{1}{\pi^2s}\delta_{ij},
\\ \nonumber
&&<\Omega'_i\Omega_j>=\frac{1}{N}\delta_{ij}(1-\frac{N}{4s}
(1-\frac{4}{\pi^2}+c_0-c_2)), \ \ \
<\underline{\Omega}_i\Omega_j>=\frac{1}{N}\delta_{ij}(1-\frac{N}{4s}),
\\ \nonumber
&&<\Omega'_i\underline{\Omega}_j>=\frac{1}{N}\delta_{ij}(1-\frac{N}{4s}
(1+c_0-c_2)),                              \hskip 2.cm
<\Omega_i\Omega_j>=\frac{1}{N}\delta_{ij},
\end{eqnarray}
This calculation was performed with formulae similar to
(\ref{2.16},\ref{2.20},\ref{2.21}), but the expression
(\ref{3.12}) was used for the Green
function $\hat{G}$, and summation over $\omega$ was restricted by
$N_{\tau}$ terms (\ref{B2}). Notice, that the presence of the member
$1-c_{\omega}$ in the numerator of the Green function (\ref{3.12}) leads
to the difference of the underlined and not underlined averages. This
difference is caused by the contribution of large $\omega \sim
\pi/\Delta$. In particularly the average $<{\bf M}^2_{tr}>$ calculated
in (\ref{2.17}) in fact coincides with the average
$<\underline{\bf M}_{tr}\cdot{\bf M}_{tr}>$ but the actual value of
$<{\bf M}^2_{tr}>$ is different and differs from the result of (\ref{2.17})
on the constant $(N-1)/4s$.

\subsection{Free energy}
\label{sec:fren}

 After the formulation of the theory for the finite time step $\Delta$ one can
calculate the GPF and the free energy of the QAF in the paramagnetic phase.
We can perform the calculation in the basic $1/2s$ approximation.
The free energy has three contributions as it follows from Eq. (\ref{2.15})
for the GPF $Z$
\begin{equation}\label{3.14}
F_{AF}=-T\ln(Z)=F_{\Omega M}+F_{\lambda}+F_{gaug}.
\end{equation}
In the lowest approximation in $1/2s$, $Z_{\Omega M}$, $Z_{\lambda}$, and
$Z_{gaug}$  are powers of determinants. The explicit form of these
determinants follows from (\ref{3.7},\ref{3.11},\ref{3.8}), and
(\ref{2.25},\ref{2.26})
\begin{eqnarray} \label{3.15}
&&F_{\Omega M}=\frac{TNN_s}{2}\sum_{\omega {\bf k}}\ln[\bar{L}(q)], \ \ \
F_{gaug}=-\frac{TN_s}{2}\sum_{\omega {\bf k}}\ln[2s\bar{Q}(q)],
\\ \nonumber
&&F_{\lambda}=\frac{TN_s}{2}\sum_{\omega {\bf k}}\ln[s^2\Pi_0(q)],
\end{eqnarray}
where all notation is in (\ref{3.11}).
Let us consider these three free energies separately. One can check
that $F_{\Omega M}$ has finite limit at $\Delta\rightarrow 0, \,
\Delta N_{\tau}=\beta$. $F_{gaug}$ and $F_{\lambda}$ do not have
finite limit at $\Delta\rightarrow 0, \, \Delta N_{\tau}=\beta$ separately,
but their sum has a finite limit.

Consider at first $F_{\lambda}$. We present it in a form
$F_{\lambda}=F_{\lambda l}+F_{\lambda h}$:
\begin{eqnarray} \label{3.16}
F_{\lambda l}=\frac{TN_s}{2}\sum_{\omega {\bf k}}
\ln[\Pi_0(q)/\Pi_{\infty}(q)], \ \ \
F_{\lambda h}= \frac{TN_s}{2}\sum_{\omega {\bf k}}
\ln\left[s^2\Pi_{\infty}(q)\right],
\end{eqnarray}
where $\Pi_{\infty}(q)=4G^{\Omega}(q)$ is the polarization operator
at large frequencies $\omega \gg T$ (\ref{2.28}) with the Green function
$G^{\Omega}(q)$ from Eq. (\ref{3.12}).
The summation over $\omega$ in the Eq. (\ref{3.16}) for $F_{\lambda l}$
is convergent because the function $\ln[\Pi_0(q)/\Pi_{\infty}(q)]$
tends to zero at large frequencies $\omega$ as $1/\omega^2$. Therefor
the summation over $\omega$ can be extended to infinity in the limit
$\Delta \rightarrow 0$.
 It is reasonable to joint the free energy $F_{\lambda h}$ with $F_{gaug}$
\begin{eqnarray} \label{3.17}
\delta F= F_{\lambda h}+F_{gaug}=\frac{TN_s}{2}\sum_{\omega {\bf k}}
\ln[2sG^{\Omega}(q)/\bar{Q}(q))]=-\frac{TN_s}{2}\sum_{\omega {\bf k}}
\ln[\bar{L}(q)]=-\frac{1}{N}F_{\Omega M}
\end{eqnarray}
The summation over the frequencies $\omega$ for $F_{\Omega M}$ can be
performed on the basis of the Appendix (\ref{sec:ApC}) and we have
the total free energy
\begin{eqnarray} \label{3.18}
F_{AF}=((N-1)/N)F_{\Omega M}+F_{\lambda l},
\end{eqnarray}
where $F_{\Omega M}$ and $F_{\lambda l}$ are
\begin{eqnarray} \label{3.19}
&&F_{\Omega M}= -NN_s{\cal J}/2+
NN_s\sum_{\bf k}\left\{\omega_{0{\bf k}}/2+
T\ln[1-\exp(-\omega_{0{\bf k}}/T)]\right\},
\\ \nonumber
&&F_{\lambda l}=\frac{TN_s}{2}\sum_{\omega {\bf k}}
\ln\left[\frac{s(\omega^2+\omega^2_{0{\bf k}})\Pi_0(q)}
{2{\cal J}(1+\gamma_{\bf k})}\right],
\end{eqnarray}
where the polarization operator $\Pi_0(q)$ is defined in (\ref{2.27}).
In this formula we neglect a small contribution of the order
$\mu^2/{\cal J}^2$.

 The first term in the free energy $F_{AF}$ (\ref{3.18}) represents the
free energy of the ordered antiferromagnet, which consists of the ground
state energy and the free energy of the magnon gas with two degenerate
degrees of freedom.

The temperature dependent part of the free energy (\ref{3.19}) at small
temperatures $T\ll {\cal J}$ is proportional to
$F_{AF}\approx N_sT^3/{\cal J}$.
Such contribution has two origins: one from $F_{\Omega M}$ and another
one from $F_{\lambda l}$.

\subsection{Perturbative corrections}
\label{sec:pert}

In this section we present the result of the calculation of corrections to
the mass operators of the $\mbox{\boldmath{$\Omega$}}$ and ${\bf M}$
fields.  The detailed analysis of such corrections is far beyond the scope
of this paper. We restrict ourselves only the lowest order of the
perturbation theory in $1/2s$. In this case these corrections can be
presented as renormalization of the initial quadratic Lagrangian
(\ref{3.7}).  It is necessary to have the Lagrangian ${\cal L}_{mod}$ and
the Hamiltonian ${\cal H}$ up to fourth order in the field ${\bf M}$,
and the Lagrangian ${\cal L}_{pha}$ up to third order.

 From the Eq. (\ref{3.3}) for ${\cal L}_{mod}$  we have
\begin{eqnarray} \label{3.20}
&&\Delta{\cal L}^{(4)}_{mod}=s\left\{2K^2-({\bf M}^2)^2-
{\bf M}^2\underline{\bf M}^2+2({\bf M}\cdot\underline{\bf M})^2\right.+
\\ \nonumber
&&\left.2K({\bf M}^2+\underline{\bf M}^2-2{\bf M}\cdot\underline{\bf M})
-2(\mbox{\boldmath{$\Omega$}}\cdot\underline{\bf M}-
\underline{\mbox{\boldmath{$\Omega$}}}\cdot{\bf M})^2\right\}/8, \ \ \
K=1-\mbox{\boldmath{$\Omega$}}\cdot\underline{\mbox{\boldmath{$\Omega$}}}.
\end{eqnarray}

 From the Eq. (\ref{3.4}) for ${\cal L}_{pha}$  we have
\begin{eqnarray} \label{3.21}
&&\Delta{\cal L}^{(3)}_{pha}=isU[
(\mbox{\boldmath{$\Omega$}}\cdot\underline{\bf M})\underline{\bf M}^2 -
(\underline{\mbox{\boldmath{$\Omega$}}}\cdot{\bf M}){\bf M}^2 -
2U(1-K)(\mbox{\boldmath{$\Omega$}}\cdot\underline{\bf M}-
\underline{\mbox{\boldmath{$\Omega$}}}\cdot{\bf M}) \cdot
\\ \nonumber
&&\cdot({\bf M}^2+\underline{\bf M}^2-2({\bf M}\cdot\underline{\bf M}))]/2 +
2isU^3(\mbox{\boldmath{$\Omega$}}
\cdot\underline{\bf M}-\underline{\mbox{\boldmath{$\Omega$}}}
\cdot{\bf M})^3/8, \ \ \
U=(1+\mbox{\boldmath{$\Omega$}}\cdot
\underline{\mbox{\boldmath{$\Omega$}}})^{-1}.
\end{eqnarray}

 The expression for the Hamiltonian ${\cal H}$ (\ref{3.5}) depends on the
scalar products of the fields $\mbox{\boldmath{$\Omega$}}, \
\underline{\mbox{\boldmath{$\Omega$}}}, \ \mbox{\boldmath{$\Omega$}}', \
\underline{\mbox{\boldmath{$\Omega$}}}', \ {\bf M}, \ \underline{\bf M}, \
{\bf M}', \ \underline{\bf M}'$. The four order over ${\bf M}$
it is rather cumbersome and we will not present its explicit
form. We only note that it is a pure algebraic problem. It is sufficient
to substitute in Eq. (\ref{3.6}) for the matrix element of the spin
operator $\hat{\bf S}$ the expression (\ref{2.5}) for the vectors ${\bf
n}_{a,b}$. After that the result must be substituted into the
expression for the Hamiltonian ${\cal H}$ (\ref{3.5}). After
expanding this expression over the field
${\bf M}$ up to fourth order we get ${\cal H}_{(4)}$.

\begin{figure}
\centering
\epsfig{figure=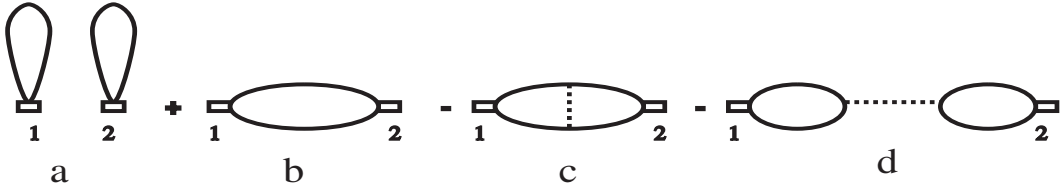,height=2.5cm,width=14.cm}
\caption{Feynmann diagrams for the averaging of four $\Omega$ fields.
The solid lines correspond the Green functions $G^{\Omega}$ of the
$\Omega$ field, the dotted lines correspond to the Green functions
of the $\lambda$ field.
The small boxes are vertices which joint two $\Omega$ fields in some
manner. In this sense the diagram $a$ describes the Gaussian contribution,
the diagram $b$ describes a correlation, the diagrams $c$ and $d$ describe
corrections due to the interaction with the $\lambda$ field.
The diagram $c$ is a small correction to the diagram $b$, but the diagram
$d$ is of the same order of magnitude as the diagram $b$.
 For the case of averaging
$<\mbox{\boldmath{$\Omega$}}^2_1\mbox{\boldmath{$\Omega$}}^2_2>=1$
the diagram $d$ simply cancels the diagram $b$. Notice, that the diagrams
$b$ and $d$ are of the order $1/N$ with respect to the diagram $a$.}
\label{fig1}
\end{figure}

On this step one can perform the averaging of the Lagrangians
${\cal L}^{(4)}_{mod}$, ${\cal L}^{(3)}_{pha}$, and the Hamiltonian
${\cal H}_{(4)}$ over the fields $\mbox{\boldmath{$\Omega$}}$ and
${\bf M}$ according to the rules (\ref{3.13}). In this point we notice that
averaging the fourth and higher powers of the $\mbox{\boldmath{$\Omega$}}$
field is a little more sophisticated. It is necessary to take into account
the fluctuation of the $\lambda$ field in the skeleton approximation over
it as it shown on Fig. \ref{fig1}.

To avoid this complication we have presented
the result in $1/N$ approximation where this complication is not
essential.

The effective kinetic Lagrangian and the Hamiltonian in the first $1/2s$
approximation are
\begin{eqnarray} \label{3.22}
&&\Delta{\cal L}_{kin}=s[a_0(1-\mbox{\boldmath{$\Omega$}}\cdot
\underline{\mbox{\boldmath{$\Omega$}}})+b_0({\bf M}^2-
{\bf M}\cdot\underline{\bf M})-ie_0(\mbox{\boldmath{$\Omega$}}
\cdot\underline{\bf M}-\underline{\mbox{\boldmath{$\Omega$}}}\cdot{\bf M})],
\\ \nonumber
&&{\cal H}=Js^2\sum_{l'\in<l>}[a_1(1-
\mbox{\boldmath{$\Omega$}}\cdot
\underline{\mbox{\boldmath{$\Omega$}}})+a_2(1-
\mbox{\boldmath{$\Omega$}}\cdot
\mbox{\boldmath{$\Omega$}}')+a_3(1-\mbox{\boldmath{$\Omega$}}'\cdot
\underline{\mbox{\boldmath{$\Omega$}}})+b_1{\bf M}^2+
\\ \nonumber
&&+b_2{\bf M}\cdot\underline{\bf M}+b_3{\bf M}\cdot{\bf M}'+b_4{\bf M}'
\cdot\underline{\bf M}-
ie_1(\mbox{\boldmath{$\Omega$}}\cdot\underline{\bf M}-
\underline{\mbox{\boldmath{$\Omega$}}}\cdot{\bf M})-
ie_2(\mbox{\boldmath{$\Omega$}}'\cdot\underline{\bf M}-
\underline{\mbox{\boldmath{$\Omega$}}}\cdot{\bf M}')],
\end{eqnarray}
where the notation is the same as in (\ref{3.7}),
and the constants $a_0,...,e_2$ are
\begin{eqnarray} \label{3.23}
&&a_i=a_i^0+g\alpha_i,\ \ \ b_i=b_i^0+g\beta_i,\ \ \ e_i=e_i^0+g\gamma_i,
\\ \nonumber
&&a_0^0=1,\ \ b_0^0=1,\ \ e_0^0=1,\ \ a_1^0=-1,\ \ a_2^0=1,\ \ a_3^0=0,\ \
\\ \nonumber
&&b_1^0=0,\ \ b_2^0=1,\ \ b_3^0=1,\ \ b_4^0=0,\ \ e_1^0=-1,\ \ e_2^0=0,\ \
\\ \nonumber
&&\alpha_0=2, \ \ \ \beta_0=(3-c_0+c_1)/2, \ \ \ \gamma_0=(7-c_0+c_1)/4,
\\ \nonumber
&&\alpha_1=2(1+c_1-c_2), \ \ \ \alpha_2=3+8/\pi^2-3c_0+2c_1+c_2,
\\ \nonumber
&&\alpha_3=-4/\pi^2+2c_0-2c_1-c_2,\ \ \ \beta_1=(5-12/\pi^2+c_1-c_2)/2,
\\ \nonumber
&&\beta_2=(-4+8/\pi^2+c_0-5c_1+4c_2)/2,\ \ \
\beta _3=(2-24/\pi^2-3c_0+c_1+2c_2)/2,
\\ \nonumber
&&\beta_4=3/\pi^2,\ \ \ \gamma_1=(-6+16/\pi^2+9c_0-c_1-8c_2)/4,\ \ \
\gamma_2=1,
\end{eqnarray}
where $g=(N-1)/4s$, $i:=0,1,2,3$. The reason why the number of the
components $N$ enters in the effective coupling constant as $N-1$ is as
follows. The short range fluctuations are directed perpendicular to
the long wave fluctuations and their number of independent components
is $N-1$.

Now one can write out the effective quadratic form for the Lagrangian
(\ref{3.22}) in the $\omega, {\bf k}$ representation. Its matrix elements
according to the notation (\ref{3.11}) are
\begin{eqnarray} \label{3.24}
&&\Lambda^{\Omega}(q)=(1-c_{\omega})[a_0+\Delta{\cal J}
(a_1+a_3\gamma_{\bf k})] +\Delta{\cal J}a_{23}(1-\gamma_{\bf k})+
\Delta\mu^2/2{\cal J}, \\
\nonumber
&&\Lambda^{M}(q)=(1-c_{\omega})[b_0-\Delta{\cal
J}(b_2+b_4\gamma_{\bf k})] +\Delta{\cal J}(b_{12}+b_{34}\gamma_{\bf k}),
\\ \nonumber
&&\Lambda^{u}(q)=s_{\omega}[e_0+\Delta{\cal J}(e_1+e_2\gamma_{\bf k})],
\ \ \ \Lambda^{d}(q)=-\Lambda^{u}(q),
\end{eqnarray}
where $a_{ij}=a_i+a_j$ and so on. The Green function is obviously
represented by Eq. (\ref{3.12}) with $\bar{L}(q)$ in a form
\begin{eqnarray} \label{3.25}
&&\bar{L}(q)=(e_0^2+\Delta{\cal J}x_1)[2(1-c_{\omega})+(1-c_{\omega})^2
\Delta{\cal J}x_2+\Delta^2\omega^2_{\bf k}],
\\ \nonumber
&&\omega^2_{\bf k}={\cal J}^2[e_0^{-2}a_{23}(1-\gamma_{\bf k})+
\mu^2/({\cal J}^2b_{1234})](b_{12}+b_{34}\gamma_{\bf k})
\end{eqnarray}
where $b_{1234}=b_{12}+b_{34}$, and the explicit form of the coefficients
$x_1, \ x_2$ follows from (\ref{3.24}). From
(\ref{2.5},\ref{3.12},\ref{3.24},\ref{3.25}) one can find the spin
correlation function in this order in $1/2s$. For that it is necessary
to take into account nonlinear corrections to the Eq.  (\ref{2.31}) which
follows from (\ref{2.5}) and also corrections to the Green functions in
the framework of this formula.

 We shall give the explicit result for the correlation radius in this
order in $1/2s$ on the basis of Eq. (\ref{2.21}). The contribution of
different frequencies $\omega$ and momenta ${\bf k}$ in this constraint
relation can be separated into two parts. The first part is the high frequency
and momentum part. To calculation this contribution it is sufficient to
take the Green function $G^{\Omega}(q)$ in the bare approximation
(\ref{3.12}) because this contribution is of the order $1/2s$.
The second contribution which is proportional to the distribution function
$n_{\bf k}$ can be considered in the continuum approximation but with
$1/2s$ corrections taken into account:
\begin{eqnarray} \label{3.26}
G^{\Omega}(q)\simeq\frac{1}{2a^2\Delta}\frac{\chi_{\perp}^{-1}}
{\omega^2+\omega^2_{\bf k}},\ \ \
\chi_{\perp}=\tilde{\rho}_s/c_s^2, \ \ \
\omega^2_{\bf k}=c^2_s{\bf k}^2+\mu^2,
\end{eqnarray}
where $\tilde{\rho}_s=Js^2a_{23}$, and
$c^2_s=e^{-2}_0a_{23}b_{1234}{\cal J}^2a^2/z$. Now, instead of the Eq.
(\ref{2.22}) we have
\begin{eqnarray} \label{3.27}
(N/4s\tilde{\rho}_s)\sum_{\bf k}\frac{n_{\bf k}}{\omega_{\bf k}}=R,\ \ \
R=1-g(1+c_0+c_1),
\end{eqnarray}
The factor $R$ inclyudes in itself the direct short wave
renormalizations. Performing the integration in the same manner as in
(\ref{2.23}) we have
\begin{eqnarray} \label{3.28}
\mu = T\exp\left(-\frac{2\pi \rho_s}{TN}\right), \ \ \
\rho_s=\tilde{\rho}_sR, \ \ \ \xi = \hbar c_{s}/\mu.
\end{eqnarray}
The actual temperature dependence is changed in the pre exponent factor
($T\rightarrow {\cal J}$) if
we take into account the long wave fluctuations in the next order in $1/N$
approximation \cite{Sac1}.

\section{Separation of scales: description of long--wave fluctuations}
\label{sec:separ}

 In some situations one can separate the long wave and short wave
spin fluctuations. Our consideration of this problem has a qualitative
character and we only sketch the possible approach to it.

The fluctuations of the $M$ field are short wave and the Green functions
$G^{M}(q),\ G^{u,d}(q)$ are regular in the long wave limit.
Therefore the separation of scales is actual for the $\Omega$ field
or the Green function $G^{\Omega}(q)$. This dressed the Green function
$G^{\Omega}(q)$ determines the action of the long--wave sigma model.
This action is universal and can easily be obtained
from the SR Lagrangian (\ref{2.37}) or from the Green function
(\ref{3.12}) by the naive long--wave limit \cite{Sac2}:
\begin{eqnarray} \label{4.1}
A=\int {\cal L}_{cont} d\tau d^2r, \ \ \
{\cal L}_{cont} = \frac{\chi_{\perp}}{2}
\left(\dot{{\bf n}}^2 +
c_s^2(\partial_i{\bf n}
\partial_i{\bf n}) + \mu ^2{\bf n}^2
\right) + i\lambda\left({\bf n}^2-1\right),
\end{eqnarray}
where $\partial_i$ is the space derivative for $i:=x,y$; $c_s$ is the
velocity of sound: $c_s^2=2J^2S^2za^2$; $\chi_{\perp}=1/2Jza^2$ is
the transverse susceptibility;
$\mu^2$ is the mass of the $n$--field in a
disordered phase\cite{Pol1}. Because the characteristic
low--energy scale $\max(T,\mu)$
is much less than the exchange constant $J,$ the long--wave
AF fluctuations contain many universal properties  \cite{Sac1}.
The connection of the constants which determine these universal properties
with parameters of the original Heisenberg model was obtained not by
a direct manner but by some calculations in the ordered phase
\cite{Iga1,Has1}.

 Our basic idea is to separate scales in the Lagrangian
${\cal L}_{tot}$ (\ref{3.2}). The simplest idea
is to separate all perturbation theory integrals over ${\bf k}$
into two parts by some separation scale $\Lambda$. But this method
introduces some arbitrariness.
We will use the scale separation based on some sort
of Pauli-Villars transformation.
The following identity holds:
\begin{eqnarray} \label{4.2}
&&I = \left(\det(\hat{G})\right)^{-1/2}\int_{-\infty}^{\infty}
\frac{d\varphi}{(2\pi)^{1/2}}
\exp \left(-\frac{1}{2}\varphi\hat{G}^{-1}\varphi - V(\varphi)\right) =
\\ \nonumber
&&\left(\det(\hat{G}_1\hat{G}_2)\right)^{-1/2}\int_{-\infty}^{\infty}
\frac{d\varphi_1d\varphi_2}{2\pi}
\exp \left(-\frac{1}{2}\varphi_1\hat{G}_1^{-1}\varphi_1
-\frac{1}{2}\varphi_2\hat{G}_2^{-1}\varphi_2 - V(\varphi_1+\varphi_2)\right),
\end{eqnarray}
where $\hat{G}=\hat{G}_1+\hat{G}_2$, and
$V(\varphi)$ is an arbitrary function such that the integrals in (\ref{4.2})
exist. This identity can be easily proved if we introduce for the second
integral new variables $\varphi=\varphi_1+\varphi_2$ and
$\psi=\varphi_1-\varphi_2$.  The integration over $\psi$ is Gaussian and
can be easily performed. As a result we have the first integral.
The identity (\ref{4.2}) may be interpreted as follows.
Let us choose
$\hat{G}=p^{-2}$ and $\hat{G}_1=(p^2(1+p^2/\Lambda^2))^{-1}, \
\hat{G}_2= (p^2+\Lambda^2)^{-1}$.
We have achieved the splitting  of the Green
function of a scalar field $\varphi$ 
into two parts with the help of the
Pauli-Villars transformation. In fact this is a strict integral identity.

 Let us apply this method to the Lagrangian ${\cal L}_{tot}$
(\ref{3.2}). We represent the field
$\mbox{\boldmath{$\Omega$}}$
as a sum two new fields ${\bf n}$ and ${\bf v}$:
$\mbox{\boldmath{$\Omega$}} = {\bf n}+{\bf v}$. The operators
$\hat{G}\equiv \hat{G}^{\Omega}(q)$, $\hat{G}_1\equiv \hat{G}^{n}(q)$,
and $\hat{G}_2\equiv \hat{G}^{v}(q)$,
in this case in the $(\omega,k)$ representation
may be chosen in the form (\ref{3.11},\ref{3.12}):
\begin{eqnarray} \label{4.3}
&&(G^{\Omega}(q))^{-1}= 2s\bar{L}(q)/\Lambda^{M}(q)\equiv
F(q)\tilde{L}(q),
\\ \nonumber
&&\tilde{L}(q)=2(1-c_{\omega})+\Delta^2 \omega^2_{0{\bf k}}, \ \ \
F(q)=2s/(1-c_{\omega}+ \Delta{\cal J}(1+\gamma_{\bf k})),
\\ \nonumber
&&(G^{n}(q))^{-1}=F(q)\tilde{L}(q)(1+\tilde{L}(q)/\Lambda^2), \ \ \
(G^{v}(q))^{-1}=F(q)(\tilde{L}(q)+\Lambda^2)
\end{eqnarray}
where the primary dispersion law $\omega_{0{\bf k}}$ is defined in
(\ref{2.15b}). We omit in (\ref{4.3}) some small terms of the order
$\Delta$ which can be essential only at calculation of the primary
free energy (\ref{sec:fren}).
We assume that $\max(T,\mu) \ll \Lambda \ll {\cal J}$. Only in this case
the cut-off momentum $\Lambda$ has a clear meaning.

The fluctuations of the field ${\bf n}$ are long wave and the low
frequency.  This is motivated by the rapid fall of the Green function
$G^{n}(q)$ with increase of the three dimensional momentum
$q=(\omega,c_s{\bf k}), \ q^2=\omega^2+c_s^2{\bf k}^2$.
As a result the Lagrangian ${\cal L}_{\Omega\lambda}$
reduces to ${\cal L}_{nv\lambda}$ and has the form:
\begin{eqnarray} \label{4.4}
&&{\cal L}_{nv\lambda} = (s/2{\cal J})
\left({\bf n}\cdot (\hat{q}^2+\mu^2)(1+\hat{q}^2/\Lambda^2){\bf n}\right)+
\\ \nonumber
&&(1/2){\bf v}\cdot F(\hat{q})(\tilde{L}(\hat{q})+\Lambda^2){\bf v}+
i\lambda({\bf n}+{\bf v})^2,
\end{eqnarray}
The long wave and the low frequency fluctuations of the field ${\bf v}$
are suppressed due to its big mass $\Lambda$, and the long wave and only
low frequency fluctuations of the fields $\Omega$ and $\lambda$ are
essential. One can check that the long wave theory for the fields
$\Omega$ and $\lambda$ practically does not contain ultraviolet
divergencies due to the cut off $\Lambda$. Actually the Green function of
the $\lambda$ field is determined by the polarization operator
$\Pi_0(q)$ (\ref{2.28}) and in the continuum limit has a form
\begin{eqnarray} \label{4.5}
G^{\lambda}(q)=(\Pi_0(q))^{-1}=sq^2/4{\cal J}, \ \ \ q \ge k_T.
\end{eqnarray}
Now we the have following large momentum behavior of the elements of the
diagram technique $G^n(q) \sim q^{-4}$, $G^{\lambda}(q) \sim q^2$, and
$\Gamma(q) \sim q^0$, where $\Gamma(q)$ is the vertex. The only divergent
diagram is the diagram for the mass operator in the lowest order but it is
naturally subtracted one time \cite{Pol1} and after that it is convergent.
An additional order of perturbation theory leads to: (a) two additional
Green functions $G^n(q)$, (b) additional Green function $G^{\lambda}(q)$
(c) additional vertex $\Gamma(q)$,  (d) additional integration over q.
As a result we have the renormalization factor
$R(q)=(G^n(q))^2G^{\lambda}(q)\Gamma(q)q^3 \sim q^{-3}$
and the general convergence is improved. This means that the
Pauli - Villars regularization is working. It is necessary to stress that
without regularization factor $\Lambda^2/(q^2+\Lambda^2)$ in the ${\bf n}$
field Green function we have $R \sim q$. This means that the original
long wave and low frequency theory is unrenormalizable, and it is not
possible to include all divergences in the finite number of objects of
the theory. Of course, the parameter $\Lambda$ is an artificial one and
must be canceled when we calculate any observable properties due
the compensation of the dependence on $\Lambda$ from long wave and short
wave contributions.

 Let us demonstrate how it is working in some basic example. Let us
consider the most important constraint of the theory which determines its
phase state: $<\mbox{\boldmath{$\Omega$}}^2>=1$. Substituting
$\mbox{\boldmath{$\Omega$}} = {\bf n}+{\bf v}$ we have
\begin{eqnarray} \label{4.6}
&&<{\bf n}^2> = 1 - <{\bf v}^2>, \ \ \
<{\bf n}^2>=N\sum_{q}G^n(q)\equiv NT\sum_{\omega=2\pi jT}
\sum_{\bf k} G^n(q) \ \ \
\\ \nonumber
&&<{\bf v}^2>=N\sum_{p}G_v (\omega,{\bf k})
=N\lim_{\Delta\rightarrow 0}\int_{-\pi/\Delta}^{\pi/\Delta}
\frac{d\omega}{2\pi}
\sum_{\bf k} G_v (q),
\\ \nonumber
&&G^n(q) = \frac{{\cal J}\Lambda^2}{s(q^2+\mu^2)(q^2+\Lambda^2)}, \ \ \
G^v(\omega,{\bf k})=\frac{\Delta^{-1}(1-c_{\omega})+
{\cal J}(1+\gamma_{\bf k})}
{2s[2\Delta^{-2}(1-c_{\omega})+\omega^2_{0{\bf k}}+\Lambda^2]}.
\end{eqnarray}
Here the summation over $j$ for the Green function $G^n(q)$ is performed
in the limits $\pm \infty$.
The main contributions in the integrals in (\ref{4.6}) from the Green
function $G^n$ are from momentum ${\bf k} \le \Lambda$, and from
the Green function $G^v$ are from momentum ${\bf k} \ge \Lambda$. This
property is general for all integrals with $G^n,\ G^v$.
The left hand side of Eq. (\ref{4.6}) can be calculated if we reformulate
accurately the summation over the frequencies $\omega = 2\pi j$ and the integration
over the momentum ${\bf k}$ can be extended
to infinity (see \cite{Man1,Sac1}).
When we calculate the right hand side of Eq. (\ref{4.6}) we can put
the temperature $T$ equal to zero and replace
the summation over $\omega$ by an
integration and the integration over $\omega$ can be easily performed.
The integration over ${\bf k}$ can be performed
treating $\Lambda/{\cal J} \ll 1$ as a small parameter.
In the part of the integral depending on
$\Lambda$, the integration over ${\bf k}$ can be
extended to infinity. The
other part is independent of $\Lambda$ and can be calculated
in the same manner as the
constant $c_{M0}$ in Eq. (\ref{2.17}). As a result, the constraint
(\ref{4.6}) has the form
\begin{eqnarray} \label{4.7}
\frac{N}{4\pi Js^2}(\Lambda+2T\ln(\mu/T)) =
1+\frac{N\Lambda}{4\pi Js^2} - \frac{N}{4s}(1+c_0-c_1). \ \ \
\end{eqnarray}
We can see that
dependence on $\Lambda$ is canceled in both
sides of Eq.  (\ref{4.6}) and for $\mu$
we have the expression (\ref{2.24}) if we take into account the main
order contribution in $1/2s$.

 Eq. (\ref{4.7}) demonstrates some general properties of the QAF
with separation of scales $\xi \ll a$. The role of the classical and
quantum spin fluctuations is essentially different. The separation of the
spin fluctuation into quantum and classical ones in perturbation
theory is determined by the summation over the frequencies $\omega$. The
Sommerfeld - Watson transformation gives us the characteristic
contributions proportional to $1+2n({\bf k})$, where $n({\bf k})$ is the
Plank distribution function, and determines the separation of
fluctuations.  The quantum fluctuations are independent on the temperature
$T$. From the point of view of the long wave theory the quantum
fluctuations are divergent in the ultraviolet region. The main
contribution into the quantum fluctuations comes from the region in the
momentum space $ka \sim 1$.  The contribution of the order $k \sim
\Lambda$ are small. Actually the second term in the right hand side of
(\ref{4.7}) is less than the last one by the small parameter
$\hbar c_s\Lambda/{\cal J}$.
This illustrate the idea that the quantum fluctuations in the long
wave region are not essential and can be neglected. In this long wave
region only the classical fluctuations with the Plank distribution
function are essential.  In this way we arrive to the "renormalized
classical region" picture \cite{Cha1,Sac1} for the classical fluctuations.
The classical long wave fluctuations possesses some nice properties.

1) They are convergent in the ultraviolet region due to a natural cut off
at $k \sim k_T$ due to the Plank functions

2) The parameters which determine these fluctuations are the
renormalized $\rho_s$, $c_s^2$,  $\chi_{\perp}=\rho_s/c_s^2$ that entered
in the action (\ref{4.1})

After this discussion the formula (\ref{3.28}) for the mass $\mu$ of the
$\Omega$ field became obvious: all the quantum renormalizations are
included in the renormalized spin stiffness $\rho_s$ and simple loop
calculation similar to (\ref{3.26},\ref{3.27}) gives this result.

The interaction between the long wave ${\bf n}$ and the $\lambda$ fields
continue to be essential which leads to the modification of the
simple result (\ref{3.27}) \cite{Cha1,Sac1}.

\section{Discussion}
\label{sec:disc}

 Our approach to the theory of QAF is based on the
explicitly rotationally invariant formulation of the pass integral in
terms of the spin coherent states. A nontrivial choice of variables
permits to formulate the theory based on the saddle point approximation
for the field $\mbox{\boldmath{$\Omega$}}$, ${\bf M}$, and
$\lambda$. As a result we can find the spin fluctuations at short
and at long distances, and at short and at long times.

We work out the method of construction of the perturbation
theory over this saddle point. It is not trivial because demands an
accurate limit to zero over the time step $\Delta$.
This permits to make the theory in some sense similar to the spin
wave theory for the calculation of the short distance and short time
fluctuations, and similar to the sigma model when we consider the long
distance and long time fluctuations.  As a result we
calculated the free energy and the first corrections to the Green
functions of the theory.

We performed the separate scales with the help of Pauli--Villars
transformation. This permits to separate out the quantum and the thermal
fluctuations. The quantum fluctuations are short waves and the thermal
fluctuations are long waves. The fluctuations of the Lagrange multiplier
$\lambda$ contains as the short wave as well as long wave
contributions.

We believe that the present approach to the QAF will be fruitful
in the theory of the AF fluctuations in HTSC superconductors.

\vskip 0.5cm  \noindent
{\Large \bf Acknowledgments}

We are grateful to A.V. Chubukov, I.V. Kolokolov, and S.Sachdev  for
stimulating discussions; C. Providencia and V.R. Vieira for the accompany
discussions; A.M. Finkelstein, and P. Woelfle  for critical remarks.  One
of the authors (V.B.) is grateful for A.L. Chernyshev, L.V. Popovich, and
V.A. Shubin for the discussion and cooperation on an earlier step of this
work.
This work was supported in part by the Portuguese projects
PRAXIS/2/2.1/FIS/451/94, V.  B. was supported in part by the Portuguese
program PRAXIS XXI /BCC/ 11952 / 97, and in part by the Russian Foundation
for Fundamental Researches, Grant No 97-02-18546.

\appendix
\section{Invariant coherent states for the rotational group}
\label{sec:apA}
\subsection{Basic formulae for the partition function in terms of the spin
coherent states}
\label{sec:apA1}

The partition function for a spin(\ref{2.2a}) system can be
presented as a functional integral over the spin coherent states
\cite{Kla1,Vie1}
\begin{eqnarray} \label{A1}
&&Z=\lim_{\Delta \rightarrow 0} \int_{-\infty}^{\infty}\cdots
\int_{-\infty}^{\infty} \prod_{j=0}^{N_{\tau}-1}\frac{(2s+1)dz'_jdz''_j}
{\pi (1+|z_j|^2)^2} \exp\left(A(z)\right),
\\ \label{A1a}
&&A(z)=-\Delta\sum_{j=0}^{N_{\tau}-1}\left[{\cal L}_{kin}(j,z)+{\cal H}(j,z)\right],
\\ \label{A1b}
&&\Delta{\cal L}_{kin}(j,z)=-2s\ln\left(\frac{1+z^*_{j+1}z_j}
{1+|z_j|^2}\right), \ \ \
{\cal H}(z)=\frac{<z^*_{j+1}|\hat{H}|z_j>}{<z^*_{j+1}|z_j>}.
\end{eqnarray}
Here $\Delta$ is the time step in the imaginary time $\tau = \Delta j$,
$N_{\tau}$ is the number of the time steps: $\Delta N_{\tau} = \beta$; $z_j$
is the complex variable for
the time step $j$; the integration is performed over $z'=\mbox{Re}(z)$ and
$z''=\mbox{Im}(z)$. It is supposed that the periodic boundary conditions
are satisfied $z_{N_{\tau}}=z_0$; $A(z)$ is the action, ${\cal L}(\tau)$ is the
Lagrangian, ${\cal H}(z)$ is the Hamiltonian.
The vector $|z>$ is the spin coherent state which is
determined in terms of the reference state $|ref> = |ss>$
\begin{eqnarray} \label{A2}
|z>=\left(1+|z|^2\right)^{-s}\exp(z\hat{S}_-)|ref>, \ \ \
\hat{S}_z|sm>=m|sm>,
\end{eqnarray}
The coherent state (\ref{A2}) is determined up to the phase factor
$\exp(i\Phi (z))$, where $\Phi (z)$ is an arbitrary function $z$, which has
no influence on the partition function $Z$, but the kinetic term
${\cal L}_{kin}(j,z)$ is changed as the time component of the Abelian gauge
field by the introduction of this phase factor
\begin{eqnarray} \label{A3}
{\cal L}_{kin}(j,z)\Rightarrow{\cal L}_{kin}(j,z)-i\Phi(z_j)+i\Phi(z_{j+1}).
\end{eqnarray}

 The complex variable $z$ has the geometric interpretation of stereographic
projection of a sphere of the radius $1/2$
\begin{eqnarray} \label{A4}
&&z= \tan(\theta/2)\exp(i\varphi), \ \ \ |z|^2=\frac{1-\cos\theta}
{1+\cos\theta},
\\ \nonumber
&& \cos\theta =\frac{1-|z|^2}{1+|z|^2}, \ \ \
\sin\theta =\frac{2|z|}{1+|z|^2}, \ \ \
\tan(\varphi)=\mbox{Im}(z)/\mbox{Re}(z),
\end{eqnarray}
where $\theta, \varphi$ are the polar and azimuthal angles which determine
the unit vector ${\bf n}=\left(\cos\varphi\sin\theta,
\sin\varphi\sin\theta, \cos\theta\right)$ which specifies the position
on the sphere of radius $1/2$.

In the continuum limit $z_{j+1}\simeq z_j + \Delta \dot{z}_j$, where the
point denotes the derivative with respect to time $\tau$,  and the
kinetic part of the Lagrangian is equal to
\begin{eqnarray}
\label{A5}
{\cal L}_{kin}(j,z)=2s\frac{\dot{z}^*z}{1+|z|^2} =
-s(1-\cos\theta)\dot{\varphi}.
\end{eqnarray}
Besides that, in the continuum limit the Hamiltonian (\ref{A1}) is
determined by the matrix element of the spin operator $\hat{\bf S}$
\begin{eqnarray} \label{A6}
<z|\hat{\bf S}|z>=s{\bf n}.
\end{eqnarray}
>From expressions (\ref{A5},\ref{A6}) Eq.(\ref{2.3a}) for
${\cal L}_{kin}(\tau,l)$ and ${\cal H}(\tau,l)$ easily follows.

\subsection{Transformation properties of the spin coherent states}
\label{sec:Tpscs}

 Further we want to clarify the reason why for the problem which
possesses obvious rotational symmetry for the Heisenberg Hamiltonian
(\ref{2.1}) the Lagrangian (\ref{2.3a}) is not explicitly rotationally
invariant.  The reason lays in the transformation properties of the state
$|z>$ under rotations. For further discussion is more convenient to use
the state $|{\bf n}>$ which coincides with the state $|z>$ up to a phase
factor \cite{Kla1,Vie1}
\begin{eqnarray} \label{A7}
|{\bf n}>=\exp(is\varphi)|z>=\exp(-i\varphi\hat{S}_z)
\exp(-i\theta\hat{S}_y)|ref>,
\end{eqnarray}
in Eq.(\ref{A7}) is supposed the complex number $z$ and the
unit vector ${\bf n}$ are connected by Eq.(\ref{A4}).  One can easily
check that the state $|{\bf n}>$ satisfies the equation
\begin{eqnarray} \label{A8}
(\hat{\bf S}\cdot{\bf n})|{\bf n}>=s|{\bf n}>.
\end{eqnarray}
Eq.(\ref{A7}) can be considered as solution of Eq.(\ref{A8}).
Eq. (\ref{A8}) is explicitly rotational invariant but its solution
is not. This circumstance was clarified by Perelomov in his
paper in which the coherent states where introduced for an
arbitrary Lee grope \cite{Per1}. The rotational grope has three
parameters: Euler angles $\varphi, \theta, \psi$ but point on the
sphere is parameterized by two parameters $\varphi, \theta$  only.
Due to that there is an invariance up to the phase factor
\begin{eqnarray} \label{A9}
|{\bf n}'>=\exp(i\Phi({\bf n},\hat{a})\hat{U}(\hat{a})|{\bf n}>, \ \
\mbox{for} \ \ n_i=a_{ij}n_j.
\end{eqnarray}
Here $\hat{a}$ is the $3\times 3$ orthogonal matrix of the three dimensional
rotations; $\Phi({\bf n},\hat{a})$ is the phase function which depends on
the vector ${\bf n}$ and the matrix $\hat{a}$; $\hat{U}(\hat{a})$ is the operator
(matrix) of the $2s+1$ dimensional representation of the rotation $\hat{a}$.
 Eq. (\ref{A9}) clarifies the situation with an invariance over
the three dimensional rotations. The vector $|{\bf n}>$ is not an
invariant object but the projector $|{\bf n}><{\bf n}|$ is an invariant
object. Now it is obvious why the total action (\ref{A1a}) and the
partition function (\ref{A1}) are invariant over rotations. Notice that
an invariance of the measure of the integration over $z$ is obvious if we
pass to variable ${\bf n}$. But the kinetic part of the Lagrangian
${\cal L}_{kin}(j,{\bf n})$ is not invariant over rotations
\begin{eqnarray} \label{A10}
\Delta{\cal L}_{kin}(j,{\bf n})\Rightarrow\Delta{\cal L}_{kin}(j,{\bf n}) -
i\Phi({\bf n_j},\hat{a})+i\Phi({\bf n_{j+1}},\hat{a}).
\end{eqnarray}
The transformation (\ref{A10}) is some kind of the Abelian lattice gauge
field transformation. It is similar to Eq.(\ref{A3}) but has different
meaning.

>From this discussion it is obvious that the module of the scalar product
of two states $|{\bf n}>$ and $|\underline{\bf n}>$ is an invariant over
rotations
\begin{eqnarray} \label{A11}
|<\underline{\bf n}|{\bf n}>|=\left(\frac{1+
\underline{\bf n}\cdot{\bf n}}{2}\right)^s.
\end{eqnarray}
Nontrivial invariant over rotation scalar, vector, tensor is determined
by the ratio
\begin{eqnarray} \label{A12}
A_{ij,...}=
\frac{<\underline{\bf n}|\hat{A}_{ij,...}|{\bf n}>}{<\underline{\bf n}
| {\bf n}>},
\end{eqnarray}
where $\hat{A}_{ij,...}$ is some operator acting in the spin
Hilbert space with definite transformation properties. When the rotational
transformation is performed the additional phase factor from numerator
and denominator cancel each other. The invariant vector one can get if we
choose instead the operator $\hat{A}_{ij,...}$ the spin operator ${\bf
S}$, see Eq.  (\ref{3.6}).

\subsection{Invariant coherent states}
\label{sec:Ics}

It seems that a noninvariance over rotations of the spin coherent
states $|{\bf n}>$ is their an integral part.
However, it is possible to define formally invariant states if we introduce
some additional unit vector ${\bf m}, \ {\bf m}^2=1$ which is orthogonal
to the vector ${\bf n}, {\bf n}\cdot{\bf m}=0$. After that it is natural to
introduce the third unit vector ${\bf k}=[{\bf n}\times{\bf m}]$. This
three unit vectors determine a reference frame. One can determine
the reference values of these unit vectors or the initial reference frame:
${\bf m}_0=(1,0,0),\ {\bf k}_0=(0,1,0),\ {\bf n}_0=(0,0,1)$. Instead of the
transformation (\ref{A7}) which define the state $|{\bf n}>$ we define
the state $|{\bf n};{\bf m}>$ with the help of the general rotation from
the reference state
\begin{eqnarray} \label{A13}
|{\bf n};{\bf m}>=\exp(-i\varphi\hat{S}_z)\exp(-i\theta\hat{S}_y).
\exp(-i\psi\hat{S}_z)|ref>
\end{eqnarray}
Due to our choice of the reference state $|ref>=|ss>$ the last
operator exponent factor in (\ref{A13}) in fact is a numerical factor.
One can easily find that $\tan\psi =-k_z/m_z$. For that it sufficient
to find three dimensional matrix which rotates the vectors
${\bf n}_0,{\bf m}_0$ into the vectors ${\bf n},{\bf m}$.

 Now one can find the Lagrangian ${\cal L}_{kin}(j,{\bf n})$
which is simply equal to
\begin{eqnarray} \label{A14}
&&\Delta{\cal L}_{kin}(j,{\bf n})=-\ln\left(
<\underline{\bf n};\underline{\bf m}|{\bf n};{\bf m}>\right),
\end{eqnarray}
where we use a general notation $\underline{x}=x(j+1),\ x\equiv x(j)$.
It is follows from Eq.(\ref{A11}) that
\begin{eqnarray} \label{A15}
&&<\underline{\bf n};\underline{\bf m}|{\bf n};{\bf m}> =
\left(\frac{1+\underline{\bf n}\cdot{\bf n}}{2}\right)^s
Y(\underline{\bf n},\underline{\bf m},{\bf n},{\bf m}),
\\ \nonumber
&&{\cal L}_{kin}={\cal L}_{mod}+{\cal L}_{pha}, \ \ \
\\ \nonumber
&&\Delta{\cal L}_{mod}=-s\ln
\left(\frac{1+\underline{\bf n}\cdot{\bf n}}{2}\right), \ \ \
\Delta{\cal L}_{pha}=-\ln\left(Y(\underline{\bf n},\underline{\bf m},
{\bf n},{\bf m})\right),
\end{eqnarray}
where $|Y(\underline{\bf n},\underline{\bf m},{\bf n},{\bf m})|=1$ is
a pure phase factor. The Lagrangian ${\cal L}_{mod}$ is pure real, and
the Lagrangian ${\cal L}_{pha}$ is pure imaginary.

An explicit form of $Y$ can be found from its
in terms of the Euler angels
\begin{eqnarray} \label{A16}
Y^2=\exp(is(\underline{\psi}-\psi))<ref|\exp(i\underline{\theta}\hat{S}_y)
\exp(i(\underline{\varphi}-\varphi)\hat{S}_z)\exp(-i\theta\hat{S}_y)|ref>
/(H.C.).
\end{eqnarray}
The matrix element in (\ref{A16}) can be calculated using Wigner
d-functions \cite{Kla1,Var1} or with the help of z-representation of
the spin coherent states (\ref{A2}), \cite{Kla1,Vie1}
\begin{eqnarray} \label{A17}
&&Y=\underline{P}^*_{\psi}P_{\psi}\underline{P}^*_{\varphi}P_{\varphi}P_n,
\ \ \ P_{\varphi} = \left(\frac{n_x-in_y}{n_x+in_y}\right)^{s/2},
\\ \nonumber
&&P_{\psi}=\left(\frac{m_z+ik_z}{m_z-ik_z}\right)^{s/2}, \ \ \
P_n=\left(\frac{1+\underline{\bf n}\cdot{\bf n}+\underline{n}_z+n_z+
i[\underline{\bf n}\times {\bf n}]_z}{H.C.}\right)^s.
\end{eqnarray}
The factor $Y$ must be an invariant over rotations but this invariability
is not obvious from the form (\ref{A17}). In fact $Y$ must be a function
of the scalar products $\underline{\bf n}\cdot{\bf n}, \
\underline{\bf n}\cdot{\bf m}, \ \underline{\bf m}\cdot{\bf k}$
and so on. To find
such kind of function we shall use a following trick. Let us calculate
the factor $Y$ in a special reference frame where
${\bf m}=(1,0,0),\ {\bf k}=(0,1,0),\ {\bf n}=(0,0,1)$. In this reference
frame $P_{\psi}=P_{\varphi}=P_n=1$ and $\underline{n}_x=
\underline{\bf n}\cdot{\bf m}, \ \underline{\bf n}_y=\underline{\bf n}
\cdot{\bf k}, \
\underline{m}_z=\underline{\bf m}\cdot{\bf n}, \ \underline{k}_z=
\underline{\bf k}\cdot{\bf n}$. As a result we have the following final
expression for the Lagrangian ${\cal L}_{pha}$
\begin{eqnarray} \label{A18}
\Delta{\cal L}_{pha}=-\frac{s}{2}\ln\left(\frac{R\underline{R}^*}
{R^*\underline{R}}\right), \ \ \
R=\underline{\bf n}\cdot{\bf m}+i\underline{\bf n}\cdot{\bf k}.
\end{eqnarray}
In (\ref{A18}) and all other places we suppose that a quantity $x$ with
double underline corresponds to a quantity $x$ without underline.
Another form for the Lagrangian ${\cal L}_{pha}$ one can get if we pass to
the reference frame where
${\bf n}=(1,0,0),\ {\bf k}=(0,1,0),\ {\bf m}=(0,0,-1)$
\begin{eqnarray} \label{A19}
&&\Delta{\cal L}_{pha}=-\frac{s}{2}\ln\left(\left[
(\underline{\bf n}\cdot{\bf n}+i\underline{\bf n}\cdot{\bf k})\cdot
(\underline{\bf m}\cdot{\bf m}+i\underline{\bf k}\cdot{\bf m})
\cdot\right.\right.
\\ \nonumber
&&\left.\left. \cdot
(1+\underline{\bf n}\cdot{\bf n}-\underline{\bf n}\cdot{\bf m}
-i\underline{\bf n}\cdot{\bf k})^2\right]\cdot[H.c.]^{-1}\right).
\end{eqnarray}

 Now one can get an expression for the Lagrangian ${\cal L}_{kin}$ in the
continuous limit. The magnitude of ${\cal L}_{mod}$ is small over
$\Delta$.
Using obvious decomposition $\underline{\bf x} \simeq {\bf x}+ \Delta
\dot{\bf x}$ for ${\bf x}:={\bf n},{\bf m},{\bf k}$ we get from (\ref{A19})
(or technically more complicated from (\ref{A18})) the following
very simple expression for ${\cal L}_{pha}$
\begin{eqnarray} \label{A20}
{\cal L}_{pha}=-is{\bf n}\cdot[{\bf m}\times\dot{\bf m}].
\end{eqnarray}

 One can check that if we choose for the vector ${\bf m}$
\begin{eqnarray} \label{A21}
&&{\bf m}=\frac{{\bf e}_z - n_z{\bf n}}{\sqrt{1-n^2_z}}, \ \ \,
{\bf m}^2=1, \ \ \ {\bf n}\cdot{\bf m}=0
\\ \nonumber
&&\mbox{we get for} \ \ \
{\cal L}_{pha}=is(1-\cos\theta)\dot{\varphi},
\end{eqnarray}
where ${\bf e}_z$ is the unit vector along the axis $z$.
This result is well known for ${\cal L}_{pha}$.

\subsection{Invariant Lagrangian for quantum antiferromagnet}
\label{sec:aplagr}

The invariant Lagrangian ${\cal L}_{mod}({\bf n})$ and the Hamiltonian
${\cal H}({\bf n})$ where presented in the text (\ref{3.3},\ref{3.5})
but the form of the Lagrangian ${\cal L}_{pha}({\bf n})$ is more
complicated and will be discussed in details below.

 The situation with the Lagrangian ${\cal L}_{pha}$ is more complicated
because it is not expressed in terms of the vectors ${\bf n}$ only
(\ref{A18}) but in terms of the vectors ${\bf m}$ and ${\bf k}$ also.
These vectors can be presented in terms of the vectors
$\mbox{\boldmath{$\Omega$}}, {\bf M}$ in a form
\begin{eqnarray} \label{A25}
&&{\bf n}_{a,b} = \pm c_m\mbox{\boldmath{$\Omega $}} + s_m{\bf l}, \ \ \
{\bf m}_{a,b}=c_m{\bf l} \mp s_m\mbox{\boldmath{$\Omega$}}, \ \ \
{\bf k}_{a,b}=\pm[\mbox{\boldmath{$\Omega$}}\times{\bf l}], \ \ \
\\ \nonumber
&&c_m=\frac{1-M^2/4}{1+M^2/4}, \ \ \  s_m=\frac{M}{1+M^2/4}, \ \  \
 c_m^2+s_m^2=1,
\\ \nonumber
&&{\bf l}=-M^{-1}[\mbox{\boldmath{$\Omega $}}\times{\bf M}], \ \ \
|{\bf l}|=1, \ \ \ {\bf f}=s_m{\bf l}, \ \ \
{\bf g}=[\mbox{\boldmath{$\Omega$}}\times{\bf f}].
\end{eqnarray}
These expressions for ${\bf m}$ and ${\bf k}$ are not analytic over
the vector ${\bf M}$: they do not have definite limit independent on
a direction of vector ${\bf M}$ when it turns to zero.
But auxiliary vectors ${\bf f}$ and ${\bf g}$ have definite limit at
${\bf M}$ turns to zero.
Now we shall prove that ${\cal L}_{pha}$ summing up over
two sublattices ${\rm a}$ and ${\rm b}$ in fact is analytic over the vector
${\bf M}$.  According to (\ref{A18}) we have
\begin{eqnarray} \label{A26}
\Delta{\cal L}_{pha}=-\frac{s}{2}\ln\left(\frac{R_a\underline{R}_a^*
R_b\underline{R}_b^*}
{R_a^*\underline{R}_aR_b^*\underline{R}_b}\right).
\end{eqnarray}
Substituting Eqs. (\ref{A25}) for ${\bf n}, \ {\bf m}$
and ${\bf k}$ in the Eq. (\ref{A18}) for $R$ we get
\begin{eqnarray} \label{A27}
&&R_{a,b}=D_{a,b}+E_{a,b}, \ \ \
D_{a,b}=\underline{c}_m[\pm c_m(\underline{\mbox{\boldmath{$\Omega$}}}
\cdot{\bf l})+i(\underline{\mbox{\boldmath{$\Omega$}}}\cdot{\bf k})],
\\ \nonumber
&&E_{a,b}=-\underline{c}_ms_m(\underline{\mbox{\boldmath{$\Omega$}}}
\cdot\mbox{\boldmath{$\Omega$}})
\mp s_m(\underline{\bf f}\cdot\mbox{\boldmath{$\Omega$}})
+c_m(\underline{\bf f}\cdot{\bf l})
\pm i(\underline{\bf f}\cdot{\bf k})
\end{eqnarray}
The quantities $D_{a,b}$ and $E_{a,b}$ are not analytic over ${\bf M}$
because they contains nonanalytic vectors ${\bf l}$ and ${\bf k}$
explicitly.
However, one can introduce new quantities $T_{a,b}=D^*_{a,b}R_{a,b}$.
Using identity $D_{a}=-D_{b}^*$ one can check that the expression
(\ref{A26})  for ${\cal L}_{pha}$ will not change if we substitute
$R_{a,b}\Rightarrow T_{a,b}$. This property is valid due to compensation
of the D-multipliers from sublattices $a$ and $b$.
Now one can check that quantities $T_{a,b}$ are analytic over ${\bf M}$
\begin{eqnarray} \label{A28}
&&T_{a,b}=A_{a,b}+B_{a,b}, \ \ \
A_{a,b}=D^*_{a,b}D_{a,b}=\underline{c}_m^2[1-
(\underline{\mbox{\boldmath{$\Omega$}}}\cdot
\mbox{\boldmath{$\Omega$}})^2 -
(\underline{\mbox{\boldmath{$\Omega$}}}\cdot{\bf f})^2],
\\ \nonumber
&&B_{a,b}=D^*_{a,b}E_{a,b}=\underline{c}_m\{\pm\underline{c}_mc_m
(\underline{\mbox{\boldmath{$\Omega$}}}\cdot\mbox{\boldmath{$\Omega$}})
(\underline{\mbox{\boldmath{$\Omega$}}}\cdot{\bf f}) +
c_m(\mbox{\boldmath{$\Omega$}}\cdot\underline{\bf f})
(\underline{\mbox{\boldmath{$\Omega$}}}\cdot{\bf f}) +
\\ \nonumber
&&\hspace*{3.7cm}   \pm(\underline{\bf f}\cdot{\bf f})
(\underline{\mbox{\boldmath{$\Omega$}}}\cdot{\bf f})
\mp(\underline{\mbox{\boldmath{$\Omega$}}}\cdot{\bf f})
\pm(\underline{\mbox{\boldmath{$\Omega$}}}\cdot\mbox{\boldmath{$\Omega$}})
(\mbox{\boldmath{$\Omega$}}\cdot\underline{\bf f})-
\\ \nonumber
&&-i[\underline{c}_m
(\underline{\mbox{\boldmath{$\Omega$}}}\cdot\mbox{\boldmath{$\Omega$}})
(\underline{\mbox{\boldmath{$\Omega$}}}\cdot{\bf g})
\pm c_m(\mbox{\boldmath{$\Omega$}}\cdot\underline{\bf f})
(\underline{\mbox{\boldmath{$\Omega$}}}\cdot{\bf g}) +
c_m(\mbox{\boldmath{$\Omega$}}\cdot\underline{\bf g})]\}.
\end{eqnarray}
Actually, quantities $A_{a,b}$ and $B_{a,b}$ (\ref{A28}) depend only on
vectors $\mbox{\boldmath{$\Omega$}}, \ {\bf f}, \ {\bf g}$
and quantity $c_m$ which all are analytic over ${\bf M}$.

\section{Calculation integrals and sum over $\omega$}
\label{sec:apB}
The calculation of integrals and sums over the frequency $\omega$ when the
finiteness of the time step $\Delta$ is taken into consideration demands
some clarification.  We start discussion from the case of the temperature
$T$ equal zero and after that pass to the case of the finite temperature.

For the case of the temperature $T$ equal zero the sum over $\omega$
over some function $f(\omega)$ pass into the integral over $\omega$ over
the one dimensional $\omega$ Brillouin band inside which
$-\pi/\Delta <\omega < \pi/\Delta$ and we are interesting in the integrals
\begin{eqnarray} \label{B1}
I=\int_{-\pi/\Delta }^{\pi/\Delta } f(\omega) d\omega/2\pi.
\end{eqnarray}
In the perturbation theory the functions $f(\omega)$
is the rational function of $c_{\omega}=
\cos(\omega\Delta)$, i.e. it is a ration of two polynomials from
$c_{\omega}$. Such ratio obviously can be presented in a form
\begin{equation}\label{B1a}
f(\omega)=P(c_{\omega}) + f'(c_{\omega}), \ \ \
f'(c_{\omega})=Q(c_{\omega})/R(c_{\omega}),
\end{equation}
where $P,Q,R$ are polynomials of $c_{\omega}$, and the power of the
polynomial $Q$ is less than the power of the polynomial $R$. The polynomial
$P(c_{\omega})$ can be easily integrated over $\omega$. The rational
function $f'(c_{\omega})$ possesses
some important properties. 1) This function can be analytically
continued in the complex $\omega$ plane. 2) It have some pole
singularities in the stripe $-\pi/\Delta <\Re(\omega)< \pi/\Delta, \
-\infty<\Re(\omega)<\infty$. 3) It turns to zero exponentially in upper
$-\pi/\Delta <\Re(\omega)< \pi/\Delta, \ 0<\Im(\omega)<\infty$
or low $-\pi/\Delta <\Re(\omega)< \pi/\Delta, \ -\infty<\Im(\omega)<0$
half stripe. 4) The function $f(\omega)$ is periodic in the $\omega$
plane $f(\omega+2\pi/\Delta)=f(\omega)$.

\begin{figure}
\centering
\epsfig{figure=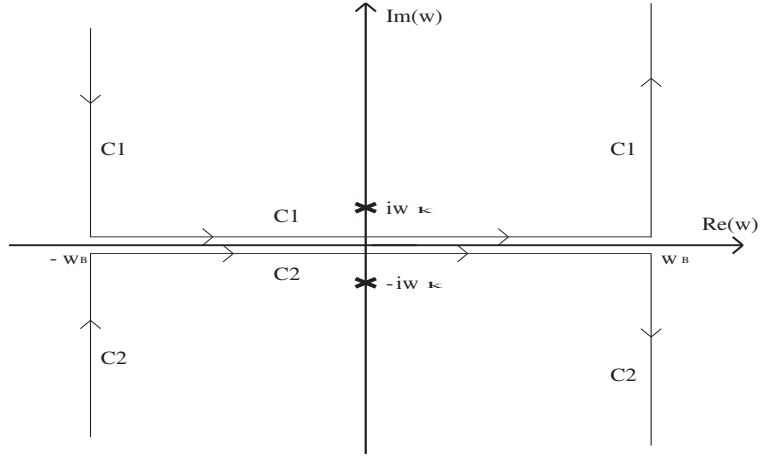,height=6.cm,width=10.cm}
\caption{Contours of integration in the complex $\omega$ plane for the
calculation of the integral from the function $f'(c_{\omega})$.
Here on the picture $w\equiv\omega$, $w_B=\pi/\Delta$, and $\pm iw_k$ are
the typical poles of a function $f'(c_{\omega})$.
The integration is performed over contour $C1$ if the function
$f'(c_{\omega})$ is decreasing at direction $\Im(\omega)>0$, and the
integration is performed over contour $C2$ if the function
$f'(c_{\omega})$ is decreasing at direction $\Im(\omega)>0$}
\label{fig2} \end{figure}

If all these conditions are valid we can convert the integral (\ref{B1})
from the function $f'(c_{\omega})$ into the contour integral in the
complex $\omega$ plane.  For that one can add to the integral (\ref{B1})
the integral over two vertical half lines which are parallel to the
imaginary $\omega$ axis.  This two lines are connects points
$(-\pi/\Delta,\pm\infty)\rightarrow (-\pi/\Delta,0)$ and
$(\pi/\Delta,0)\rightarrow(\pi/\Delta,\pm\infty)$.  The choice of the half
plane for these two half axes is dictated by the direction in which the
function $f(\omega)$ fall down exponentially.  Due to periodicity the
function $f(\omega)$ the contour integrals over $\omega$ along each pair
of this axes compensate each other. If we add these two integrals for the
function $f(\omega)$ to the primary integral (\ref{B1}), and also the
integral over the segment $(-\pi/\Delta,\pm\infty)\rightarrow
(\pi/\Delta,\pm\infty)$ (the last one must be infinitely small) we shall
have the closed integral in the $\omega$ plane. It is equal to the sum
residues of the function $f(\omega)$ in this half stripe (see Fig.
\ref{fig2}).

 For the case of the final temperature it is necessary to calculate the
following sum
\begin{eqnarray} \label{B2}
I=T\sum_{j=-n_{\tau}}^{n_{\tau}} f(\omega), \ \ \
\omega=2\pi jT, \ \ \ n_{\tau}=(N_{\tau}-1)/2,
\end{eqnarray}
where for definiteness the number of the $N_{\tau}$ summation points is
odd. Using representation (\ref{B1a}) one can easily perform the summation
over $\omega$ of the polynomial $P(c_{\omega})$ and for the second term
this sum can be converted by the standard manner into the contour integral
\begin{eqnarray} \label{B3}
I'=\frac{1}{4\pi
i}\int_{C} \cot\left(\frac{\beta\omega}{2} \right)f'(c_{\omega})d\omega,
\end{eqnarray}
where the contour of the integration is around of the
segment of the real $\omega$ axis $-\pi/\Delta <\Re(\omega)< \pi/\Delta$.
By the same manner as it was discussed for the case the temperature $T=0$
one can supplement the contour $C$ up to two closed contours over two the
half stripes $-\pi/\Delta <\Re(\omega)< \pi/\Delta,\Im\omega>0 $ and
$-\pi/\Delta <\Re(\omega)< \pi/\Delta,\Im\omega<0 $. As a result the sum
(\ref{B2}) will be equal to the sum of the residues of the function
$0.5\cot(0.5\beta\omega)f'(c_{\omega})$ in these two half stripes.

\section{Calculation of the free energy}
\label{sec:ApC} \ \

The free energy of simple oscillator is
\begin{eqnarray}   \label{C1}
F=\frac{T}{2}\sum_{\omega}\ln\left[\lambda_{\omega}\right], \ \ \
\lambda_{\omega}=2(1-c_{\omega})+\Delta^2\omega_0^2,
\end{eqnarray}
where $\omega=2\pi jT$, and $-(N_{\tau}-1)/2<j<(N_{\tau}-1)/2$,
for definiteness $N_{\tau}$ is odd.
It is convenient to represent the eigenvalues  $\lambda_{\omega}$
in the form
\begin{eqnarray}   \label{C2}
&&\lambda_{\omega} = \mu_{\omega} \mu^*_{\omega},\ \ \
\mu_{\omega} = \alpha  + \beta \exp(i\omega\Delta),
\\ \nonumber
&& (\alpha ,\beta ) \simeq \pm 1 +\Delta\omega_0/2
\end{eqnarray}
In terms of the quantities $\mu_{\omega}$ we have for $F$
\begin{eqnarray}   \label{C3}
&&F =T\sum_{\omega}\ln\left[\mu_{\omega}\right]=
TN_{\tau}\ln[\alpha] + T\sum_{\omega}\ln[1 + (\beta/\alpha)
\exp(i\omega\Delta )]
\end{eqnarray}
The sum over $\omega$ in Eq. (\ref{C3}) can be easily
performed if we expand the logarithmic function in a series over $x =
(\alpha/\beta)\exp(i\omega\Delta )$ and take into
account  only the terms $x^p$ with $p = kN_{\tau}$, where $k$ is integer,
which give nonzero contributions.
The result for $F$ has the form
\begin{eqnarray}   \label{C4}
F = TN_{\tau}\ln[\alpha] +T\ln[1 + (\beta/\alpha)^{N_{\tau}}].
\end{eqnarray}
Using Eq. (\ref{C2}) for $\alpha$, and $\beta$  and conditions
$\Delta\omega_0 \ll 1, \  \Delta N_{\tau}=1/T$ we easily get for
the free energy
\begin{eqnarray}   \label{19}
F =\omega_0/2+T\ln[1 - \exp(-\omega_0 T)].
\end{eqnarray}
It is well known expression for the free energy of the harmonic oscillator.

\end{document}